\documentclass[iop,numberedappendix]{emulateapj-rtx4}

\usepackage{txfonts}
\usepackage{multirow}
\usepackage{verbatim}
\usepackage{color}
\usepackage{sidecap}
\usepackage{natbib}

\definecolor{orange}{rgb}{1,.5,0}

\newcommand{\hd}{H$_2$}
\newcommand{\ndhp}{N$_2$H$^+$}
\newcommand{\nddp}{N$_2$D$^+$}
\newcommand{\nht}{NH$_3$}
\newcommand{\ccs}{CCS}

\newcommand{\tco}{$^{13}$CO}
\newcommand{\cdo}{C$^{18}$O}
\newcommand{\cthd}{c-C$_3$H$_2$}
\newcommand{\cthdt}{c-C$_3$HD}
\newcommand{\chtoh}{CH$_3$OH}
\newcommand{\cdh}{C$_2$H}
\newcommand{\cn}{CN}
\newcommand{\cs}{CS}
\newcommand{\cts}{C$^{34}$S}
\newcommand{\co}{CO}
\newcommand{\so}{SO}
\newcommand{\tso}{$^{34}$SO}
\newcommand{\sod}{SO$_2$}
\newcommand{\ocs}{OCS}
\newcommand{\hcn}{HCN}
\newcommand{\nhdd}{NH$_2$D}
\newcommand{\hcop}{HCO$^+$}
\newcommand{\hctn}{HC$_3$N}

\def\pipe{Pipe nebula}
\def\paper{Paper~I}
\def\paperd{Paper~II}
\def\paperud{Papers~I and II}
\def\stem{\textit{stem}}
\def\bowl{\textit{bowl}}
\def\polp{$p_\%$}
\def\deltapa{$\delta$P.A.}
\def\mi{\textit{abundance}}

\def\mis{\textit{abundances}}

\newcommand{\tenpow}[2]{#1$\times$10$^{#2}$}

\def\lsim{\rlap{\lower4pt\hbox{$\sim$}}\raise1pt\hbox{$<$}}  
\def\gsim{\rlap{\lower4pt\hbox{$\sim$}}\raise1pt\hbox{$>$}}  

\def\Tmb{\mbox{$T_{\rm MB}$}}
\def\Av{\mbox{$A_{\rm V}$}}

\def\nhd{\mbox{$n_{\rm H_2}$}}
\def\Nhd{\mbox{$N_{\rm H_2}$}}

\def\mjybeam{mJy~beam$^{-1}$}
\def\kms{\mbox{km~s$^{-1}$}}
\def\cmt{cm$^{-3}$}

\def\msun{\mbox{$M_\odot$}}

\def\asec{.$\!''$}
\def\deg{$^\circ$}

\slugcomment{}
\shorttitle{Starless Cores in the Pipe Nebula II}
\shortauthors{Frau et al.}

\begin{document}

\title{ Young starless cores embedded in the magnetically dominated Pipe
Nebula. II.\\ Extended dataset
\footnote{B\MakeLowercase{ased on observations
carried out with the} IRAM 
\MakeLowercase{30-m telescope}. IRAM
\MakeLowercase{is supported by} INSU/CNRS 
(F\MakeLowercase{rance}), MPG
(G\MakeLowercase{ermany), and} IGN
(S\MakeLowercase{pain}).}}

\author{
P.\ Frau\altaffilmark{1}, 
J.\ M.\ Girart\altaffilmark{1}, 
M.\ T.\ Beltr\'an \altaffilmark{2}, 
M.\ Padovani \altaffilmark{1},
G.\ Busquet \altaffilmark{3}, 
O.\ Morata \altaffilmark{4},\\
J.\ M.\ Masqu\'e \altaffilmark{5}, 
F.\ O.\ Alves \altaffilmark{6},
\'A.\ S\'anchez-Monge \altaffilmark{2}, 
G.\ A.\ P.\ Franco \altaffilmark{7},
and
R.\ Estalella \altaffilmark{5}
}

\affil{$^1$ Institut de Ci\`encies de l'Espai (CSIC-IEEC), Campus UAB, Facultat de Ci\`encies, Torre C-5p, 08193 Bellaterra, Catalunya, Spain}
\affil{$^2$ INAF-Osservatorio Astrofisico di Arcetri, Largo Enrico Fermi 5, 50125 Firenze, Italy}
\affil{$^3$ INAF-Istituto di Astrofisica e Planetologia Spaziali, via Fosso del Cavaliere 100, 00133 Roma, Italy}
\affil{$^4$ Institute of Astronomy and Astrophysics, Academia Sinica, P.O.\ Box 23-141, Taipei 10617, Taiwan}
\affil{$^5$ Departament d'Astronomia i Meteorologia and Institut de Ci\`encies
del Cosmos (IEEC-UB),\\Universitat de Barcelona, Mart{\'\i} i Franqu\`es 1, 08028 Barcelona, Catalunya, Spain}
\affil{$^6$ Argelander-Institut f\"ur Astronomie der Universit\"at Bonn, Auf dem H\"ugel 71, 53121 Bonn, Germany}
\affil{$^7$ Departamento de F\'isica - ICEx - UFMG, Caixa Postal 702, 30.123-970, Belo Horizonte, Brazil}

\date{\it
Received 2012 May 9;
accepted 2012 July 13;
published
}

\begin{abstract}

The \pipe\ is a massive, nearby, filamentary dark molecular cloud with a low
star-formation efficiency threaded by a uniform magnetic field perpendicular to
its main axis. It harbors more than a hundred,  mostly quiescent, very
chemically young starless cores.  The cloud is, therefore, a good laboratory to
study the earliest stages of the star-formation process. We aim to investigate
the primordial conditions and the relation among physical, chemical, and
magnetic properties in the evolution of low-mass starless cores.  We used the
IRAM 30-m telescope to map the 1.2~mm dust continuum emission of five new
starless cores, which are in good agreement  with previous visual extinction
maps. For the sample of nine cores, which  includes the four cores studied in a
previous work, we derived a \Av\ to \Nhd\ factor of
(1.27$\pm$0.12)$\times$10$^{-21}$~mag~cm$^{2}$ and a background visual
extinction of $\sim$6.7~mag possibly arising from the cloud material. We derived
an average core diameter of $\sim$0.08~pc, density of $\sim$10$^5$~\cmt,  and
mass of $\sim$1.7~\msun. Several trends seem to exist related to increasing core
density: ({\it i}) diameter seems to shrink, ({\it ii}) mass seems to increase,
and ({\it iii}) chemistry tends to be richer. No correlation is found between
the direction of the surrounding diffuse medium magnetic field and the projected
orientation of the cores, suggesting that large scale magnetic fields
seem to play a secondary role in shaping the cores. We also used the IRAM 30-m
telescope to extend the previous molecular survey at 1 and 3~mm of early- and 
late-time molecules toward the same five new \pipe\ starless cores, and analyzed
the normalized intensities of the detected molecular transitions.  We confirmed
the chemical differentiation toward the sample and increased the number of
molecular transitions of the ``diffuse'' (e.g. the ``ubiquitous'' \co, \cdh, and
\cs), ``oxo-sulfurated'' (e.g. \so\ and \chtoh), and ``deuterated'' (e.g. \ndhp,
\cn, and \hcn) starless core groups. The chemically defined core groups seem to
be related to different evolutionary stages: ``diffuse'' cores present the cloud
chemistry and are the less dense, while ``deuterated'' cores are the densest and
present a chemistry typical of evolved dense cores. ``Oxo-sulfurated'' cores
might be in a transitional stage exhibiting intermediate properties and a very
characteristic chemistry.

\end{abstract}


\section{Introduction}\label{intro}

\begin{figure*}[t]
   \centering
   \includegraphics[width=14cm,angle=0]{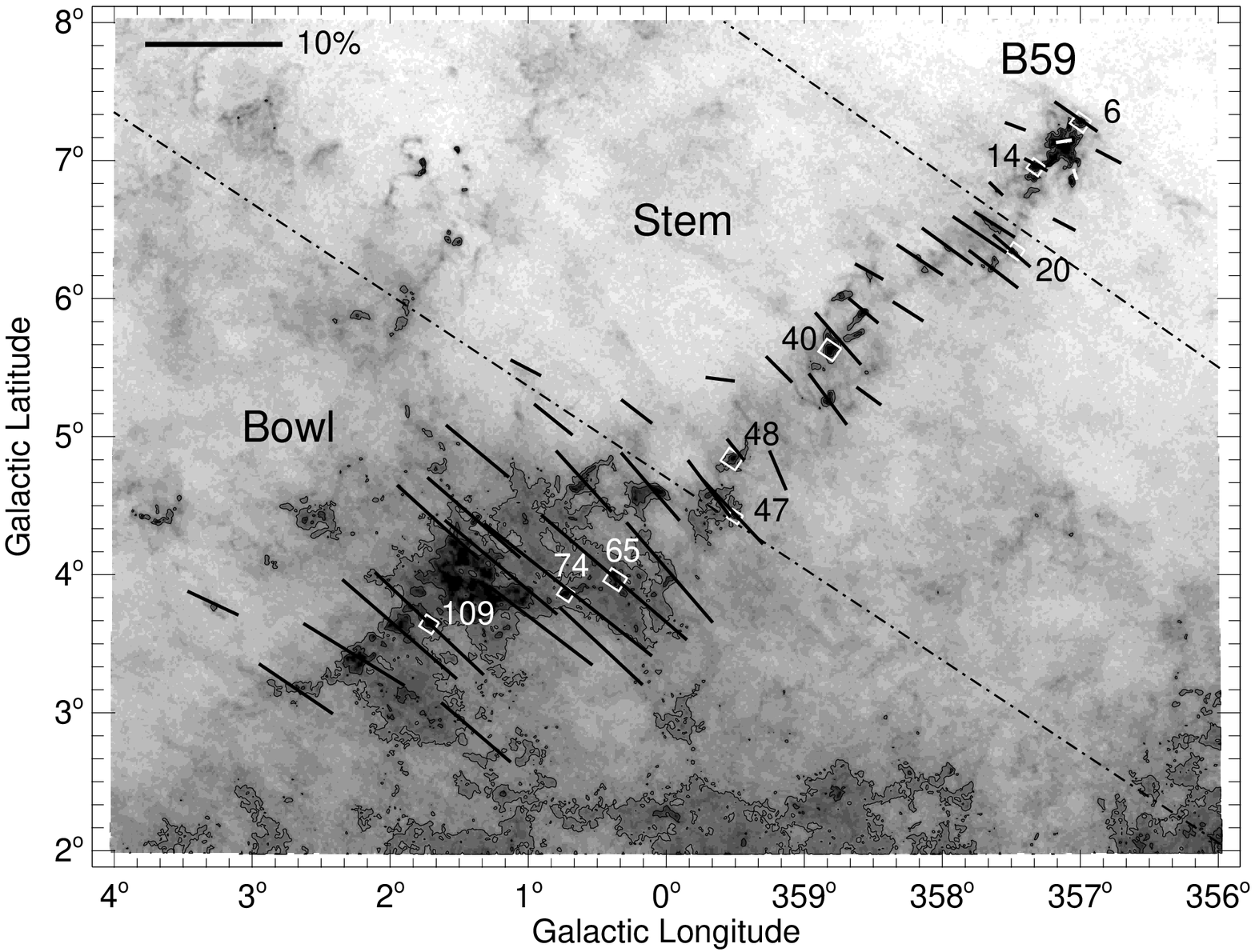}

\caption{Position of the observed cores plotted over the 2MASS extinction map of
the \pipe\ \citep{lombardi06}. Black segments represent the mean polarization
vector of the region \citep{alves08} with the scale shown on the top left corner
of the figure. White boxes depict the size of the 1.2~mm continuum maps
(Section~\ref{obs_mambo} and Fig.~\ref{fig:mambo} of both \paper\ and the
present work). The dashed lines separate the three different magnetically
defined regions \citep{alves08}. The lowest visual extinction ($A_\texttt{v}$)
corresponds  to 0.5 magnitudes. The highest $A_\texttt{v}$ is observed toward
the \bowl\ and B59 regions, where it reaches approximately 20 magnitudes. 
\label{fig:pipe} }  

    \end{figure*}

The \pipe\ is a massive ($10^4$~\msun: \citealp{onishi99}) nearby (145~pc:
\citealp{alves07}) filamentary dark cloud located in the southern sky
(Fig.~\ref{fig:pipe}). What differentiates the \pipe\ from other low-mass
star-forming regions such as Taurus and $\rho$-Ophiuchus is that it is very
quiescent and has a very low star-formation efficiency, only the Barnard~59
(B59) region shows star formation \citep{forbrich09,brooke07,roman09,roman11}. 
The cloud harbors more than one hundred low-mass starless dense cores in a very
early evolutionary stage \citep{muench07,rathborne08}. Thermal pressure appears
to be the dominant source of internal pressure of these cores: most of them
appear to be pressure confined, but gravitationally unbound \citep{lada08}.
Only the B59 region shows a significant non-thermal contribution to
molecular line widths that could be caused by outflows feedback
\citep{duartecabral12}. Through simulations of an unmagnetized cloud compatible
to the \pipe,   \citet{heitsch09} predicted pressures lower than those required
by \citet{lada08}. This result suggests that an extra source of pressure, such
as magnetic fields, is acting. In fact, \citet{franco10} found that most of the
\pipe\ is magnetically dominated and that turbulence appears to be
sub-Alfv\'enic. \citet{alves08} have distinguished three regions in the cloud
with differentiated polarization properties, proposed to be at different
evolutionary stages (Fig.~\ref{fig:pipe}). B59, with low polarization degree
(\polp) and high polarization vector dispersion (\deltapa), is the only
magnetically supercritical region and might be the most evolved, the \stem\
would be at an earlier evolutionary stage, and finally, the \bowl, with high
\polp\ and low \deltapa, would be at the earliest stage. The \pipe\ is, hence,
an excellent place to study the initial conditions of core formation which may
eventually undergo star formation.

\begin{table}[t]
\caption{
Source List Observed in \paper\ and in this Work.
}
\begin{tabular}{ccccc}
\hline\hline
& \multicolumn{1}{c}{$\alpha$(J2000)} 
& \multicolumn{1}{c}{$\delta$(J2000)} 
& \multicolumn{1}{c}{$\texttt{v}_{\rm LSR}$$^b$}
& \multicolumn{1}{c}{} \\
\multicolumn{1}{c}{Source$^a$} &
\multicolumn{1}{c}{h m s} &
\multicolumn{1}{c}{$\degr$ $\arcmin$ $\arcsec$} &
\multicolumn{1}{c}{(km s$^{-1}$)} &
\multicolumn{1}{c}{Region$^c$} \\
\hline
Core 06  & 17 10 31.6 & -27 25 51.6 & +3.4 & B59\\
Core 14  & 17 12 34.0 & -27 21 16.2 & +3.5 & B59\\
Core 20  & 17 15 11.2 & -27 35 06.0 & +3.5 & {\it Stem}\\
Core 40  & 17 21 16.4 & -26 52 56.7 & +3.3 & {\it Stem}\\
Core 47  & 17 27 29.6 & -26 59 06.0 & +2.8 & {\it Stem}\\
Core 48  & 17 25 59.0 & -26 44 11.8 & +3.6 & {\it Stem}\\
Core 65  & 17 31 20.5 & -26 30 36.1 & +5.0 & {\it Bowl}\\
Core 74  & 17 32 35.3 & -26 15 54.0 & +4.2 & {\it Bowl}\\
Core 109 & 17 35 48.5 & -25 33 05.8 & +5.8 & {\it Bowl}\\ 
\hline
\end{tabular}

$^a$ According to \citet{lombardi06} numbering. \\
$^b$ \citet{rathborne08}. \\
$^c$ According to \citet{alves08} diffuse gas polarimetric properties.
\label{tab_source}
\end{table}

\citet[][hereafter \paper]{frau10} presented the first results of a molecular
line study at high spectral resolution for a sample of four cores distributed
in the different regions of the \pipe. The aim of the project was to chemically
date the cores through an extensive molecular survey based in two main
categories of molecules: early- and late-time \citep[e.g.,][]{taylor96}. In
addition, we mapped the 1.2~mm dust continuum emission of the cores. We found
no clear correlation between the chemical evolutionary stage of the cores and
their location in the \pipe\ and, therefore, with the large scale magnetic
field. However, at core scales, there are hints of a correlation  between the
chemical evolutionary stage of the cores and the local magnetic properties.
Recently, \citet[][hereafter \paperd]{frau12} have presented a 3~mm
$\sim$15~GHz wide chemical survey toward fourteen starless cores in the \pipe. 
In order to avoid a density bias, we defined the molecular line normalized
intensity by dividing the spectra by the visual extinction (\Av) peak, similar
to the definition of the abundance. We found a clear chemical differentiation,
and normalized intensity trends among the cores related to their \Av\ peak
value. We defined three groups of cores: ``diffuse'' cores (\Av$\lsim$15~mag)
with emission only of ``ubiquitous'' molecular transitions present in all the
cores (\cdh, \cthd, \hcop, \cs, \so, and \hcn), ``oxo-sulfurated'' cores
(\Av$\simeq$15-22~mag) with emission of molecules like \tso, \sod,
and \ocs, only present in this group, and finally, ``deuterated'' cores
(\Av$\gsim$22~mag), which present emission in nitrogen- and deuterium-bearing
molecules, as well as in carbon chain molecules.

In this paper, we further explored observationally the relationship among
structure, chemistry, and magnetic field by extending the sample in five new
\pipe\ cores, for a total of nine, and several new molecular transitions. We
repeated and extended the analysis conducted in \paper\ for molecular
(temperature, opacity, and column density estimates) and continuum (dust
parameters estimates and comparison with previous maps) data. We also derived and
analyzed the molecular line normalized intensities as in \paperd. For the sake of
simplicity, we omit here technical details, which are widely explained in
\paperud.

\section{Observations and data reduction\label{obs}}

\subsection{MAMBO-II observations\label{obs_mambo}}

We mapped cores~06, 20, 47, 65, and 74 (according to \citealp{lombardi06}
numbering) at 1.2~mm with the 117-receiver MAMBO-II bolometer (array diameter
of 240$''$) of the IRAM 30-m telescope in Granada (Spain). Core positions are
listed in  Table~\ref{tab_source}. The beam size is $\sim$11\arcsec\ at 250~GHz.
The observations were carried out in March and April 2009 and in January and
March 2010, in the framework of a flexible observing pool, using the same
technique and strategy as in \paper.  A total of 16 usable maps were selected
for analysis: 4 for cores~06 and 74, 3 for cores~20 and 47, and 2 for core~65.
The weather conditions were good, with zenith optical depths between 0.1 and
0.3 for most of the time. The average corrections for pointing and focus stayed
below 3$''$ and 0.2~mm, respectively. The maps were taken at an elevation of
$\la$25$^{\circ}$ because of the declination of the sources.

The data were reduced using MOPSIC and figures were created using the GREG
package (both from the GILDAS\footnote{MOPSIC and GILDAS data reduction
packages are available at http://www.iram.fr/IRAMFR/GILDAS} software).

\begin{table}[t]
\caption{
Molecular transitions observed toward the \pipe\ cores with the IRAM 30-m antenna.
}
\scriptsize
\begin{tabular}{cc r @{.} l r @{.} l r @{/} l cc}
\hline\hline
&&\multicolumn{2}{c}{Frequency} 
&\multicolumn{2}{c}{Beam$^a$} 
&\multicolumn{2}{c}{Beam} 
&\multicolumn{1}{c}{$\Delta {\tt v}$$^c$} \\
 \multicolumn{1}{c}{Molecule} 
&\multicolumn{1}{c}{Transition} 
&\multicolumn{2}{c}{(GHz)} 
&\multicolumn{2}{c}{($''$)} 
&\multicolumn{2}{c}{efficiency$^b$} 
&\multicolumn{1}{c}{(km\,s$^{-1}$)}
&\multicolumn{1}{c}{Type$^d$}\\
\hline
C$_3$H$_2$      &(2$_{1,2}$--1$_{1,0}$)  & 85  & 3389 & 28 & 8  &0.78&0.81 &0.07  & E\\
C$_2$H		&(1--0)			 & 87  & 3169 & 28 & 1  & --\phantom{8}&0.81 &0.07  & E\\
HCN             &(1--0)                  & 88  & 6318 & 27 & 7  &0.78&0.81 &0.07  & E\\
N$_2$H$^+$      &(1--0)                  & 93  & 1762 & 26 & 4  &0.77&0.80 &0.06  & L\\
C$^{34}$S	&(2--1)			 & 96  & 4130 & 25 & 5  & --\phantom{8}&0.80 &0.06  & E\\
CH$_3$OH        &(2$_{-1,2}$--1$_{-1,1}$)& 96  & 7394 & 25 & 4  & --\phantom{8}&0.80 &0.06  & L?\\
CH$_3$OH        &(2$_{0,2}$--1$_{0,1}$)  & 96  & 7414 & 25 & 4  & --\phantom{8}&0.80 &0.06  & L?\\
CS              &(2--1)                  & 97  & 9809 & 25 & 1  &0.76&0.80 &0.06  & E\\
C$^{18}$O       &(1--0)                  & 109 & 7822 & 22 & 4  & --\phantom{8}&0.78 &0.05  & E\\
$^{13}$CO       &(1--0)                  & 110 & 2014 & 22 & 3  & --\phantom{8}&0.78 &0.05  & E\\
CN              &(1--0)                  & 113 & 4909 & 21 & 7  &0.75&0.78 &0.05  & E\\
C$^{34}$S	&(3--2)			 & 146 & 6171 & 16 & 8  & --\phantom{8}&0.74 &0.04  & E\\
CS		&(3--2)			 & 146 & 6960 & 16 & 8  & --\phantom{8}&0.73 &0.04  & E\\
N$_2$D$^+$      &(2--1)                  & 154 & 2170 & 16 & 0  &0.77&0.72 &0.04  & L\\
DCO$^+$         &(3--2)                  & 216 & 1126 & 11 & 4  &0.57&0.62 &0.03  & L\\
C$^{18}$O       &(2--1)                  & 219 & 5603 & 11 & 2  & --\phantom{8}&0.61 &0.03  & E\\
$^{13}$CO       &(2--1)                  & 220 & 3986 & 11 & 2  & --\phantom{8}&0.61 &0.05  & E\\
CN              &(2--1)                  & 226 & 8747 & 10 & 9  &0.53&0.60 &0.03  & E\\
N$_2$D$^+$      &(3--2)                  & 231 & 3216 & 10 & 6  &0.67&0.59 &0.03  & L\\
H$^{13}$CO$^+$  &(3--2)                  & 260 & 2554 &  9 & 5  &0.53&0.53 &0.02  & L\\
\hline
\end{tabular}

$^a$ [HPBW/$''$]=2460$\times$[freq/GHz]$^{-1}$\\ {\scriptsize
(http://www.iram.es/IRAMES/telescope/telescopeSummary/telescope\_summary.html)}\\
$^b$ ABCD and EMIR receiver, respectively \\
$^c$ Spectral resolution.\\
$^d$ E = Early-time. L = Late-time. See \paper\ for details.
\label{tab_obs}
\end{table}

\subsection{Line observations\label{obs_line}}

\begin{figure*}[ht]
   \centering
   \includegraphics[width=12cm,angle=-90]{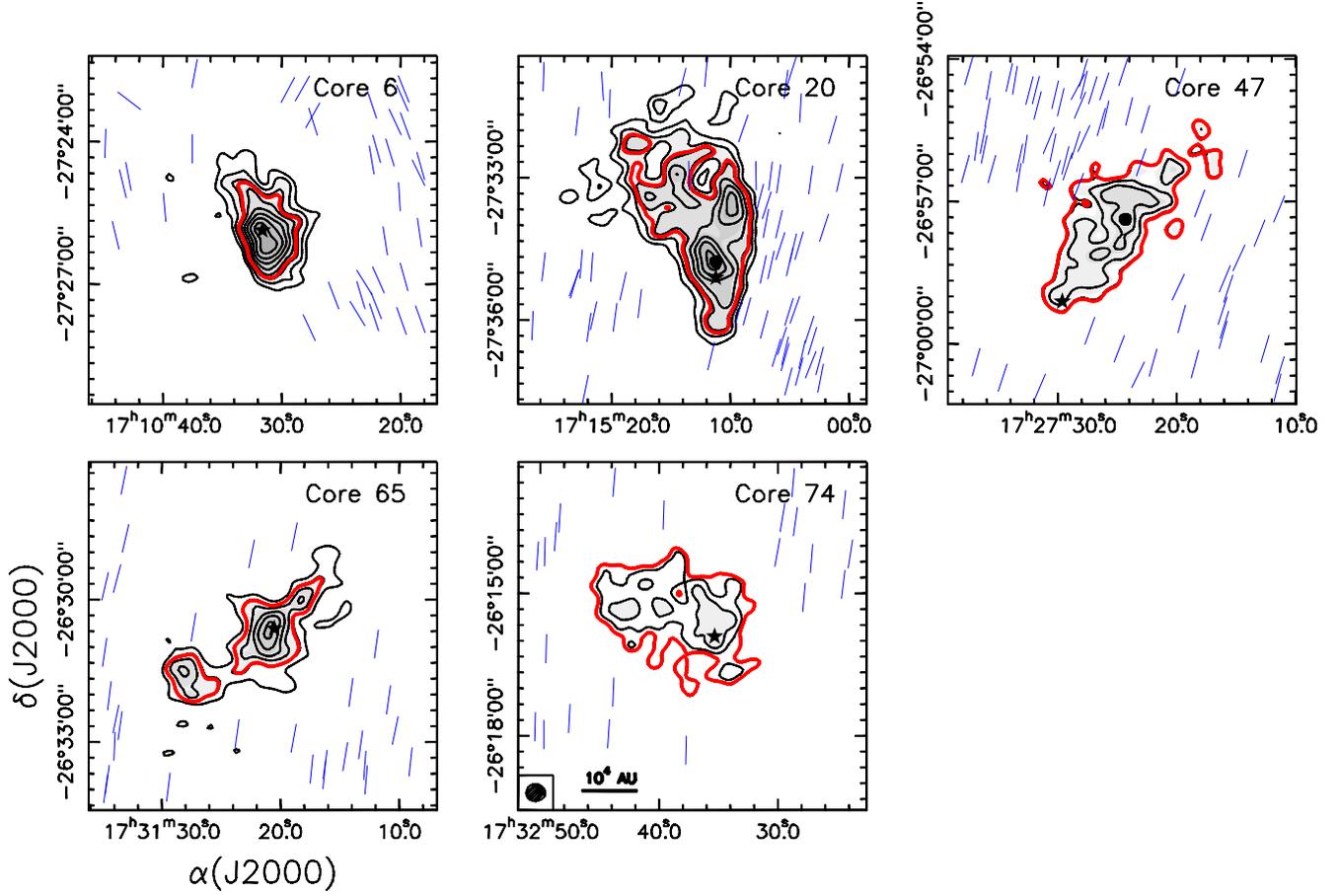}
     
\caption{IRAM 30-m MAMBO-II maps of the dust continuum emission at 1.2~mm
toward five cores of the \pipe.  The grayscale levels for all the maps are 3 to
18 times 5.75~mJy\,beam$^{-1}$. The contour levels are 3 to 10 times $\sigma$ 
in steps of 1-$\sigma$, where 1-$\sigma$ is  4.0, 4.5, 4.9, 4.4, and
4.3~mJy\,beam$^{-1}$ for core~06, 20, 47, 65, and 74, respectively. The red
thick contour marks the half maximum emission level of the source 
(Table~\ref{tab_param_mambo}). Black stars indicate the \cdo\ pointing center
reported  by \cite{muench07}. Black filled circles indicate the position 
where line observations have been performed, if different from the
\citet{muench07} position, closer to the dust continuum emission maximum which
falls into the beam area. The blue vectors depict the polarization vectors
found by \cite{franco10}.  In the bottom left corner of the bottom middle panel
the convolved beam and the spatial scale for the maps are shown.
\label{fig:mambo}}

   \end{figure*}

We performed pointed observations within the regions of the cores~06, 20, 47,
65, and 74  with the ABCD and EMIR heterodyne receivers of the IRAM 30-m
telescope covering the 3, 2, 1.3 and 1.1~mm bands. The observed positions were
either the \cdo\ pointing center reported by \citet[][depicted by star symbols
in Fig.~\ref{fig:mambo}]{muench07}, or the pointing position closer to the dust
continuum peak (circle symbols in Fig.~\ref{fig:mambo}). The
epochs, system configuration, technique, and methodology used are the same as
in \paper. We present also new molecular transitions observed toward the whole
sample of nine
cores in the same epochs as \paper: \chtoh, \tco\ and \cdo\ in the (1--0) and
(2--1) transitions, and \cs\ and \cts\ in the (3--2) transition.  System
temperatures, in \Tmb\ scale, ranged from 200 to 275~K (3~mm) and from 440 to
960~K (1~mm) for good weather conditions, and reached 450~K (3~mm) and 3200~K
(1~mm) for bad weather conditions.

Additional pointed observations were performed in August 2011 toward the dust
emission peak of cores 06, 14, 20, 40, 47, 48, 65, and 109
(Table~\ref{tab_param_mambo} of both \paper\ and the present work). The EMIR E0
receiver, together with the VESPA autocorrelator at a spectral resolution of
20~kHz, was tuned to the \cdh~(1--0) transition. Six spectral windows were set
to the six hyperfine components of the transition (spanning from 87.284 to
87.446~GHz; Table~4 of \citealp{padovani09}). Frequency switching mode was used
with a frequency throw of 7.5~MHz. System temperatures ranged from $\sim$100 to
$\sim$125~K.

Table~\ref{tab_obs} shows the transitions and frequencies observed, as well as
the beam sizes and efficiencies. Column 6 lists the velocity resolution
corresponding to the channel resolution of the VESPA autocorrelator
(20~kHz). Column 7 specifies the evolutive category of each molecule (i.e.
early- or late-time molecule). We reduced the data using the CLASS package of
the GILDAS software. We obtained the line parameters either from a Gaussian fit
or from calculating their statistical moments when the profile was not Gaussian.


\section{Results and analysis\label{sec_res}}

In this Section, we present the dust continuum emission maps for five new
\pipe\ cores to be analyzed together with the four cores already presented in
\paper. We also present molecular line observations for the new five cores in
the same transitions presented in \paper, as well as several new transitions
for the nine cores. Finally, following \paperd\ analysis, we derive the
normalized intensities of the detected molecular transitions. A detailed
explanation of the methodology can be found in \paperud.

\begin{table*}[ht]

\caption{
1.2~mm continuum emission parameters.
}
\begin{tabular}{ccccccccccc}
\hline
&
\multicolumn{1}{c}{$\alpha$(J2000)$~^a$} &
\multicolumn{1}{c}{$\delta$(J2000)$~^a$} &
\multicolumn{1}{c}{$T_{\rm dust}$} &
\multicolumn{1}{c}{rms} &
\multicolumn{1}{c}{$S_{\nu}$} &
\multicolumn{1}{c}{$I_\nu^{\textrm{\tiny{Peak}}}$} &
\multicolumn{1}{c}{Diameter $^b$} &
\multicolumn{1}{c}{$N_{\rm H_2}$ $^c$} &
\multicolumn{1}{c}{$n_{\rm H_2}$ $^c$} &
\multicolumn{1}{c}{Mass $^c$} 
 \\
\multicolumn{1}{c}{Source} &
\multicolumn{1}{c}{h m s} &
\multicolumn{1}{c}{$\degr$ $\arcmin$ $\arcsec$} &
\multicolumn{1}{c}{(K)} &
\multicolumn{1}{c}{(mJy beam$^{-1}$)} &
\multicolumn{1}{c}{(Jy)} &
\multicolumn{1}{c}{(mJy  beam$^{-1}$)} &
\multicolumn{1}{c}{(pc)} &
\multicolumn{1}{c}{(10$^{21}$cm$^{-2}$)} &
\multicolumn{1}{c}{(10$^{4}$cm$^{-3}$)} &
\multicolumn{1}{c}{($M_{\odot}$)} \\
\hline

Core 06
	&	17 10 31.8 	&	$-$27 25 51.3 	& 	10.0$^d$
& 4.0   &0.58   
& 42.6   &0.051   &16.18   $^d$&15.44   $^d$& 0.88   $^d$\\
Core 20
	&	17 15 11.5 	&	$-$27 34 47.9	& 	15.2$^e$	
& 4.5   &1.52   
& 42.6   &0.088   & 7.33   & 4.04   & 1.20   \\
Core 47
	&	 17 27 24.3	&	 $-$26 57 22.2	&	12.6$^e$ 
& 4.9   &0.73   
& 28.5   &0.093   & 4.17   & 2.18   & 0.76   \\
Core 65
	&	17 31 21.1	&	$-$26 30 42.8	&	10.0$^d$   
& 4.4   &0.48   
& 36.1   &0.053   &12.39   $^d$&11.38   $^d$& 0.73   $^d$\\
Core 74
	&	 17 32 35.3	&	$-$26 15 54.0	&	10.0$^d$  
& 4.3   &0.40   
& 21.4   &0.097   & 3.11   $^d$& 1.56   $^d$& 0.61   $^d$\\

\hline
\end{tabular}

$^a$ Dust continuum emission peak.\\
$^b$ Diameter of the circle with area equal to the source area satisfying 
$I_\nu$$>$$I_\nu^{\rm Peak}/2$\\
$^c$ Assuming $\kappa_{\rm 250~GHz}$$=$0.0066~cm$^2$~g$^{-1}$ as a medium value 
between dust grains with thin
and thick ice mantles \citep{ossenkopf94}. See Appendix~1 in \paper\ 
 for details on the calculation. \\
$^d$ No kinetic temperature estimate, therefore we assumed 10~K 
based on the average temperatures of the other cores in the 
\pipe\ \citep{rathborne08}. \\ 
$^e$ Adopted to be equal to the kinetic temperature estimated from 
\nht\ \citep{rathborne08}.\\
\label{tab_param_mambo}
\end{table*}

\subsection{Dust continuum emission\label{res_dust}}

In Fig.~\ref{fig:mambo} we present the MAMBO-II maps of the dust continuum
emission at 1.2~mm toward the five new cores of the \pipe, convolved to a
21\asec5 Gaussian beam in order to improve the signal-to-noise ratio (SNR), and
to smooth the appearance of the maps. Table~\ref{tab_param_mambo} lists the peak
position of the 1.2~mm emission after convolution, the dust temperature
\citep{rathborne08}, the rms noise of the maps, the flux density and the value
of the emission peak. Additionally, we also give the derived full width half
maximum (FWHM) equivalent diameter, \hd\ column density (\Nhd), \hd\ volume
density (\nhd) density, and mass for each core (see Appendix~A in \paper\ for
details). These parameters are derived from the emission within the 3-$\sigma$
level assuming $\kappa_{\rm 250~GHz}$$=$0.0066~cm$^2$~g$^{-1}$ as a
medium value  between dust grains with thin and thick ice mantles
\citep{ossenkopf94},  and discussed in Section~\ref{sec_discuss}. 

The flux density of the cores ranges between $\sim$0.40 and $\sim$1.52~Jy, while
the peak value ranges between $\sim$21 and $\sim$43~\mjybeam. The maps of
Fig.~\ref{fig:mambo} show the different morphology of the five cores. Core~06,
located in the most evolved B59 region, shows one of the weakest emission levels
($\sim$0.6~Jy) of the present sample. It is the most compact ($\sim$0.05~pc) and
densest ($\sim$1.5$\times$10$^5$~\cmt) core of the five. It shows similar
physical properties to core~14 (\paper), also in the B59 region.  The two cores
located in the \stem, 20 and 47, show a very diffuse nature with elliptical
morphologies similar to the previously presented \stem\ core~48. The three of
them have similar physical properties in terms of size ($\sim$0.09~pc) and
density ($\sim$3$\times$10$^4$~cm$^{-3}$). The \bowl\ cores, 65 and 74, do not
show a defined morphology. Their sizes, densities and masses are very different.
Core~65 is more compact and denser, while core~74 shows properties comparable to
those of the \stem\ cores. The morphology of the dust continuum emission for all
the cores is in good agreement with that of previous extinction maps
\citep{lombardi06,roman09}.

\begin{table}[t]
\caption{
Summary of detections and upper levels in K toward the \pipe\ cores$^a$.
}
\scriptsize
\begin{tabular}{c ccccccccc}
\hline\hline\noalign{\smallskip}
Molecular &\multicolumn{9}{c}{Core}\\
\cline{2-10}
transitions     & 06 & 14 & 20 & 40 & 47 & 48 & 65 & 74 & 109\\
\noalign{\smallskip}
\hline\noalign{\smallskip}
C$_3$H$_2$~(2$_{1,2}$--1$_{1,0}$) & $\surd$  & $\surd$ & 0.22 & $\surd$  & $\surd$ & 0.07& 0.06& 0.17 & $\surd$ \\
C$_2$H~(1--0) & $\surd$  & $\surd$  & $\surd$  & $\surd$  & $\surd$  & $\surd$ & 0.06& -- & $\surd$ \\
HCN~(1--0) & $\surd$ & 0.21 & $\surd$  & $\surd$  & $\surd~^b$  & $\surd$  & -- & 0.18 & $\surd$ \\
N$_2$H$^+$~(1--0) & $\surd$  & $\surd$ & 0.08 & $\surd$ & $\surd$& 0.07& 0.06& 0.11 & $\surd$ \\
C$^{34}$S~(2--1) & $\surd$  & $\surd$  & $\surd$  & $\surd$  & $\surd$  & $\surd$ & 0.07 & $\surd$  & $\surd$ \\
CH$_{3}$OH~(2--1) & $\surd$  & $\surd$  & $\surd$  & $\surd$  & $\surd$  & $\surd$  & $\surd$  & $\surd$  & $\surd$ \\
CS~(2--1) & $\surd$  & $\surd$  & $\surd$  & $\surd$  & $\surd~^b$  & $\surd$  & --  & $\surd$  & $\surd$ \\
C$^{18}$O~(1--0) & $\surd$  & $\surd~^b$  & $\surd~^b$  & --  & $\surd~^b$  & --  & --  & $\surd$  & $\surd~^b$ \\
$^{13}$CO~(1--0) & $\surd$  & $\surd~^b$  & $\surd~^b$  & --  & $\surd~^b$  & --  & --  & $\surd$  & $\surd~^b$ \\
CN~(1--0) & $\surd$  & $\surd$ & 0.25 & $\surd$  & $\surd$ & 0.17& 0.11& 0.19 & $\surd$ \\
C$^{34}$S~(3--2)& 0.10 & $\surd$  & $\surd$  & -- & 0.06& 0.16& 0.08& 0.24 & $\surd$ \\
CS~(3--2) & $\surd$  & $\surd$  & $\surd$  & $\surd$  & $\surd$  & $\surd$  & $\surd$  & $\surd$  & $\surd$ \\
N$_2$D$^+$~(2--1)& 0.04& 0.12& 0.17 & $\surd$ & 0.09& 0.08& 0.08& 0.09 & $\surd$ \\
DCO$^+$~(3--2) & $\surd$ & 1.71& 2.33& 0.61 & -- & 0.76 & -- & 0.50 & $\surd$ \\
C$^{18}$O~(2--1) & $\surd$  & $\surd~^b$  & $\surd~^b$  & $\surd~^b$  & $\surd~^b$  & $\surd$  & $\surd$  & $\surd$  & $\surd~^b$ \\
$^{13}$CO~(2--1) & $\surd$  & $\surd~^b$  & $\surd~^b$  & --  & $\surd~^b$  & --  & --  & $\surd$  & $\surd~^b$ \\
CN~(2--1)& 0.80& 0.97 & -- & 1.70 & -- & 0.76 & -- & 0.84& 0.90\\
N$_2$D$^+$~(3--2)& 1.00& 1.01& 1.27~$^b$& 0.93 & -- & 1.94 & -- & 1.32& 0.91\\
H$^{13}$CO$^+$~(3--2)& 1.29& 1.52& 1.94$~^b$& 1.40 & -- & 2.38 & -- & 1.84& 1.34\\
\hline
\end{tabular}

$^a$ \paper\ results are included. The transitions marked with -- have 
not been observed. Those marked with $\surd$ have been detected
 toward the corresponding core. Otherwise, the 3$\sigma$ upper 
 limit is shown. In the seventh column of Table~\ref{tab_obs},
 early- and late-time labels are listed for each molecule.\\
$^b$ Observed toward the extinction peak. \\
\label{tab_detect}
\end{table}

   \begin{figure*}[ht]
   \centering
   \includegraphics[height=\textwidth,angle=270]{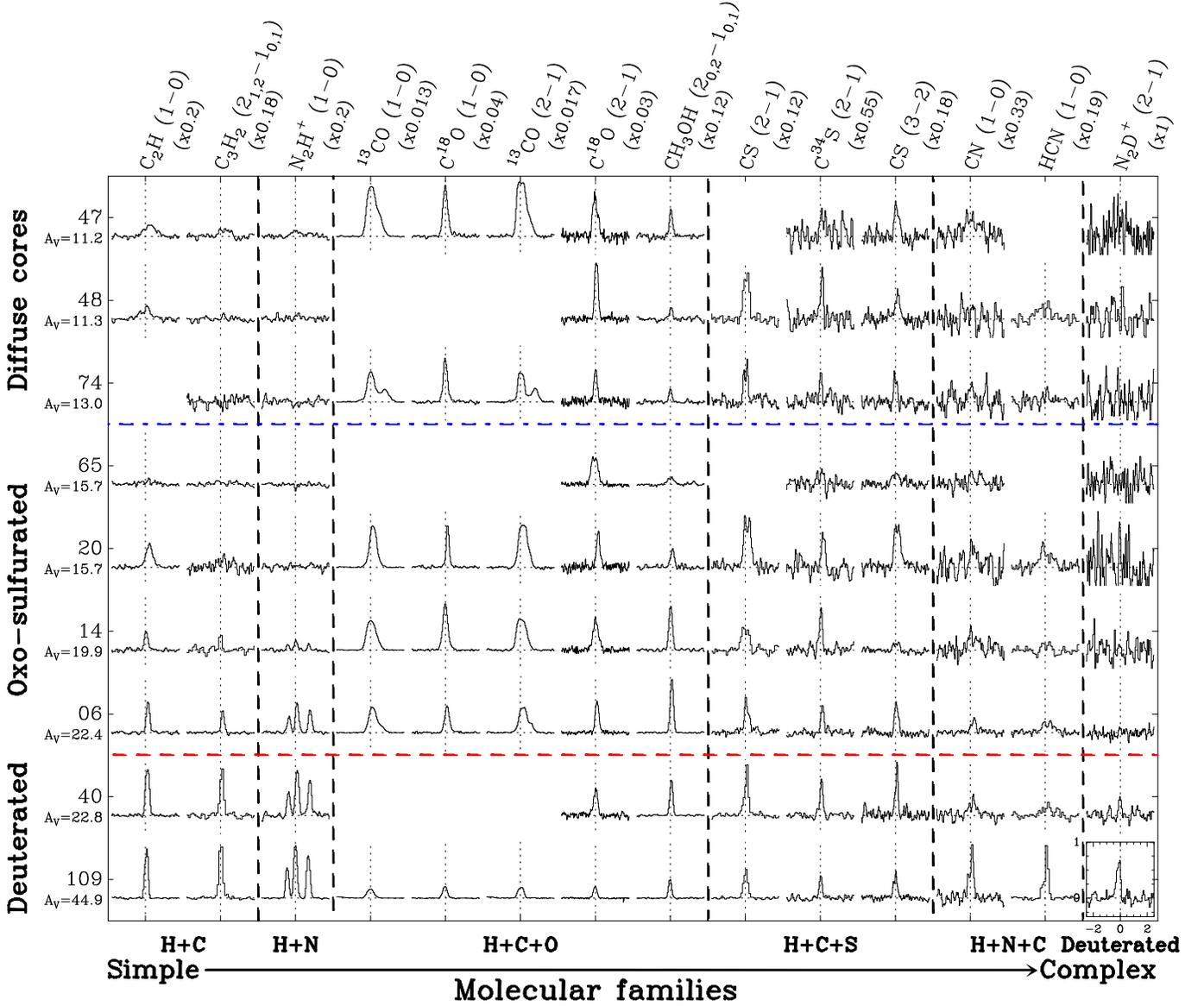}

\caption{Selected normalized molecular transitions toward the observed cores.
The brightest transition is shown for the \cdh, \chtoh, and \hcn\ lines. The
scale is shown in the bottom right spectrum. The normalized intensity axis
ranges from -0.33 to 1, while the velocity axis spans 5~\kms\ centered at the
systemic velocity of the core. {\it Rows}:  individual cores, labeled on the
left-hand side of the figure, ordered by its \Av\ peak. {\it Columns}:
molecular transition, ordered by molecular families, labeled on the top of the
figure. The spectra have been divided by $[ \Av/100 \, {\rm mag}]$ to mimic the
abundance, where the \Av\ value is that at the respective core center
\citep{roman10} given below the core name. Each molecular transition has been
multiplied by a factor, given below its name, to fit in a common scale.
\label{fig_norm}}

\end{figure*} 

   \begin{figure*}[ht]
   \centering
   \includegraphics[height=\textwidth,angle=270]{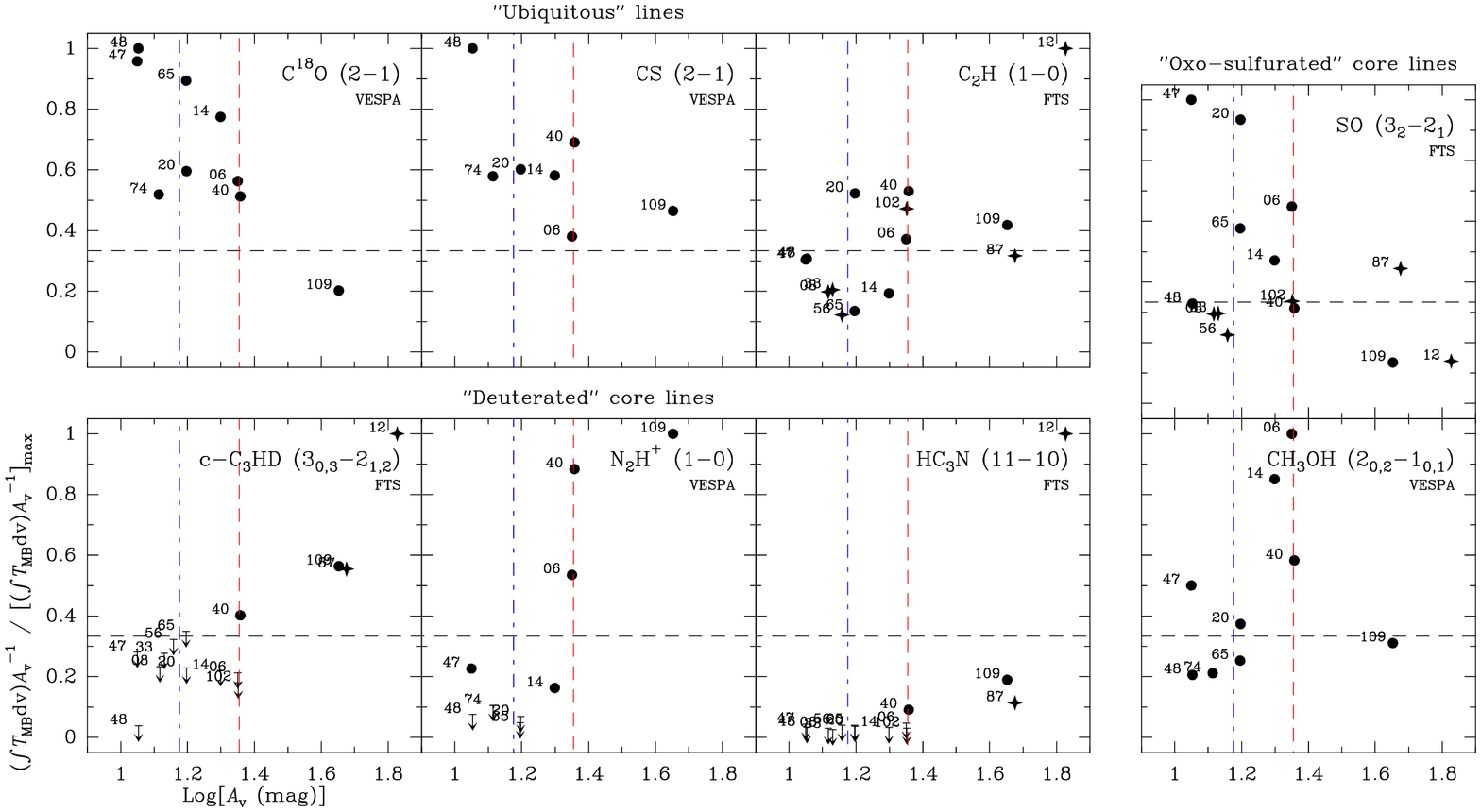}

\caption{Normalized integrated intensity (\Av$^{-1}$ $\int T_{\rm MB} dv$) of
selected molecular transitions divided by its maximum value as a function of
the logarithm of the core \Av\ peak. Each panel shows the molecular transition
in the top right corner. The backend is labeled below the molecular transition:
VESPA panels present the nine cores of \paper\ and present work, while FTS
panels present the fourteen cores of \paperd. Filled circles represent the nine
cores of \paper\ and present work, stars represent the other six cores from
\paperd\ (08, 12, 33, 56, 87, and 102), and arrows represent upper limits. 
Molecules have been split into the three categories defined in \paperd, labeled
on the top of each group. The blue dot-dashed and red dashed vertical lines
mark the transition from ``diffuse'' to ``oxo-sulfurated'', and from
``oxo-sulfurated'' to ``deuterated'' cores, respectively. The horizontal dashed
line marks a third of the peak value which helps to show the behavior change of
the ``oxo-sulfurated'' and ``deuterated'' lines in our sample. The maximum
values for each molecular transition are, in K~\kms~mag$^{-1}$, 0.114, 0.036,
and 0.030 for the ubiquitous \cdo, \cs, and \cdh, 0.059 and 0.025 for the
``oxo-sulfurated'' \so\ and \chtoh, and 0.003, 0.010, and 0.019 for the
``deuterated'' \cthdt, \ndhp, and \hctn, respectively.} \label{fig_areaNorm}

\end{figure*}

\subsection{Molecular survey of high density tracers\label{res_line}}

We present molecular line data observed toward the dust continuum emission peak
or toward the \cdo\ peak position reported by \citet[][for more details see
Fig.~\ref{fig:mambo}]{muench07}, defined as the core center and supposed to
exhibit brighter emission from molecular transitions. As discussed in \paper,
the chemical properties derived toward the dust emission peak are representative
of the chemistry at the core center. Our higher resolution dust emission maps
show a peak offset with respect to the \cdo\ for cores 20 and 47. For the former
one, the offset is only $\sim$20\arcsec\ while for the latter, more diffuse one,
the offset is of $\sim$130\arcsec (see Section~\ref{ind-cores}).

In Figs.~\ref{fig:lineshfs}--\ref{fig:lines1_new},  we show the spectra of the
different molecular transitions observed toward the dust continuum emission
peak of each core. Figures~\ref{fig:lineshfs}--\ref{fig:lines1} show the
molecular transitions with and without hyperfine components, respectively, for
the five new cores. The new molecular transitions for the whole sample of nine
cores are shown in Figs.~\ref{fig:lineshfs_new}--\ref{fig:lineshfs_new2}, for
those with hyperfine components, and Fig.~\ref{fig:lines1_new}, for those
without hyperfine components.  Table~\ref{tab_detect} summarizes the detections
or the 3$\sigma$ upper limits of the non detections toward the whole sample of
nine cores. We found that early-time molecules are broadly detected over the
whole sample. Several of them  were detected toward all the cores:
\chtoh~(2--1), \cs\ in the (2--1) and (3--2) transitions, and \tco\ and \cdo\
in the (1--0) and (2--1) transitions.  On the other hand, only a few cores
present emission of late-time molecules. The cores with $n_{\mathrm
H_2}$$\gsim$10$^5$~cm$^{-3}$ (06, 14, 40, and 109 but not 65) presented more
detections than shallower cores and, indeed, were the only cores presenting
\ndhp\ emission. Tables~\ref{tab_param1}--\ref{tab_param3} give the parameters
of the detected lines. Regarding the line properties, the $\texttt{v}_{\rm
LSR}$ measured for different species are generally consistent within 0.2~\kms. 
Intensities vary significantly over the sample: cores~06, 40 and 109 are
generally bright while the rest of the sample shows fainter emission.
``Bright'' lines ($T_{\rm MB}$$>$0.2~K) are mostly very narrow
(0.2\lsim$\Delta\texttt{v}$\lsim0.3~\kms), although transitions of \co\ and
\cs\ isotopologues can show broader profiles
($\Delta\texttt{v}$$\lsim$0.5~\kms\ if ``bright''). In some cases, this
broadening can be explained in terms of a second velocity component generally
merged with the main one (cores~06 and 20 in \cs, and cores~06, 14, 47, 74, and
109 in \tco). 

In addition to the line parameters, we derived the molecular column densities
for all the detected species (see Appendix~B in \paper\ for details) which are
listed in Tables~\ref{tab_mol_col_dens1}--\ref{tab_mol_col_dens2}. For the
transitions with detected hyperfine components (\cdh, \hcn, \ndhp, \chtoh, and
\cn), we derived the opacity using the hyperfine components fitting method (HFS)
of the CLASS package.  For the molecular transitions observed in more than one
isotopologue, this is \cs\ and \cts\ in the (2--1) and (3--2) transitions, and
\tco\ and \cdo\ in the (1--0) and (2--1) transitions, we derived numerically
the opacity. Table~\ref{tab_dust_col_dens} shows the \hd\ column density of the
cores for different angular resolutions.
Tables~\ref{tab_abun1}--\ref{tab_abun2} give the molecular abundances with
respect to \hd. 

Figure~\ref{fig_norm} shows the normalized intensities with respect to the core
\Av\ peak of a selection of detected molecular transitions toward the sample of
\paper\ and the present work. Some of the lines were already presented in
\paperd\ (except for core~74), although here are shown with a higher spectral
resolution (e.g., the 3~mm \cdh, \cthd, and \hcn\ line). The \tco\ and \cdo\
isotopologues can be considered as ``ubiquitous'' because they are present in
all the observed cores (for cores 40, 48, and 65, the CO lines present intense 
emission but were observed toward a position that  is offset from the core peak
position). Their general trend is to decrease as density  increases.  The
\cts~(2--1) line, which is optically thin, shows a similar trend as the  main
isotopologue (see \paperd), considered also as ``ubiquitous''. The decrease in 
the normalized intensity in the \cs\ lines is only apparent for the densest
core~109. The \cn\ normalized intensities are larger toward the densest cores,
which suggests that \cn\ is typically associated with the ``deuterated'' group.
Late time species, such as \ndhp\ and \nddp, are only present in the densest
cores and their emission tends to be brighter with increasing density,
confirming that both species are typical of the ``deuterated'' group. These
general results are in agreement with the observational classification of cores
presented in \paperd, which is based on a wider molecular survey at 3~mm.

Finally, we defined the observational normalized integrated intensity  (NII) as
$\left(\int T_{\rm MB} dv \right) /A_{\rm V}$, to illustrate the different
behavior of the molecular transitions that motivated the observational core
classification proposed in \paperd.  Figure~\ref{fig_areaNorm} shows NII
divided by its maximum value in the sample for selected molecular transitions
typical of the three core groups: ``diffuse'', ``oxo-sulfurated'', and
``deuterated''. ``Ubiquitous'' species are present in all the cores and their
NII tends to decrease as the central density increases indicating  possible
depletion. ``Oxo-sulfurated'' species show low NII values except for a narrow
range of densities (\Av$\simeq$15-22~mag). \chtoh~(2$_{0,2}$--1$_{0,1}$) shows
a similar behavior to the ``oxo-sulfurated'' species previously identified
(e.g. \so, \sod, and \ocs; \paperd) but seems to peak at slightly larger
densities (\Av$\sim $20--23~mag). ``Deuterated'' species are only present
toward the densest cores and their NII values increase with increasing density.

\subsection{LTE status through hyperfine structure}

\begin{figure}
\begin{center}
{\includegraphics[width=\columnwidth,angle=0]{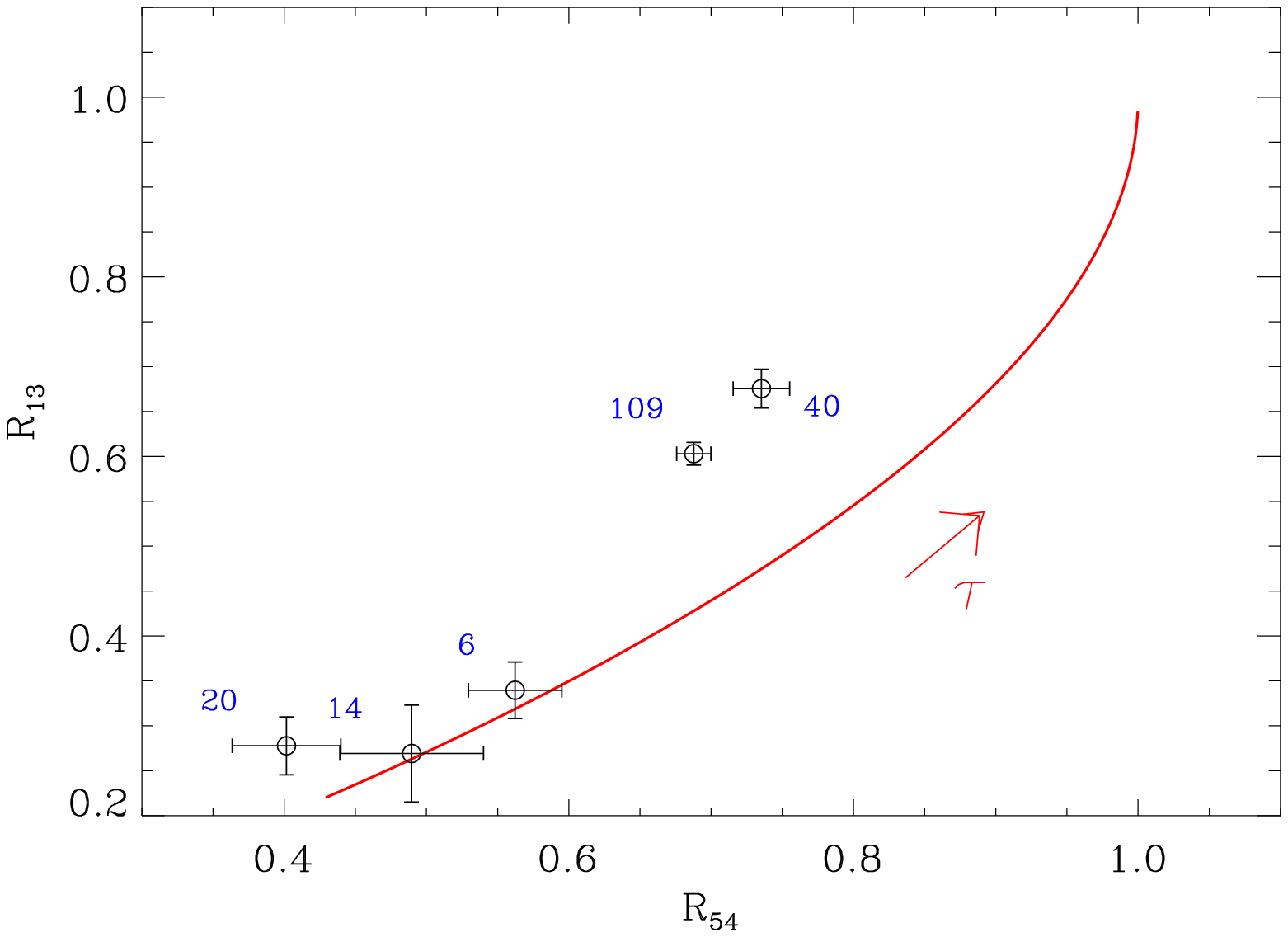}} 
{\includegraphics[width=\columnwidth,angle=0]{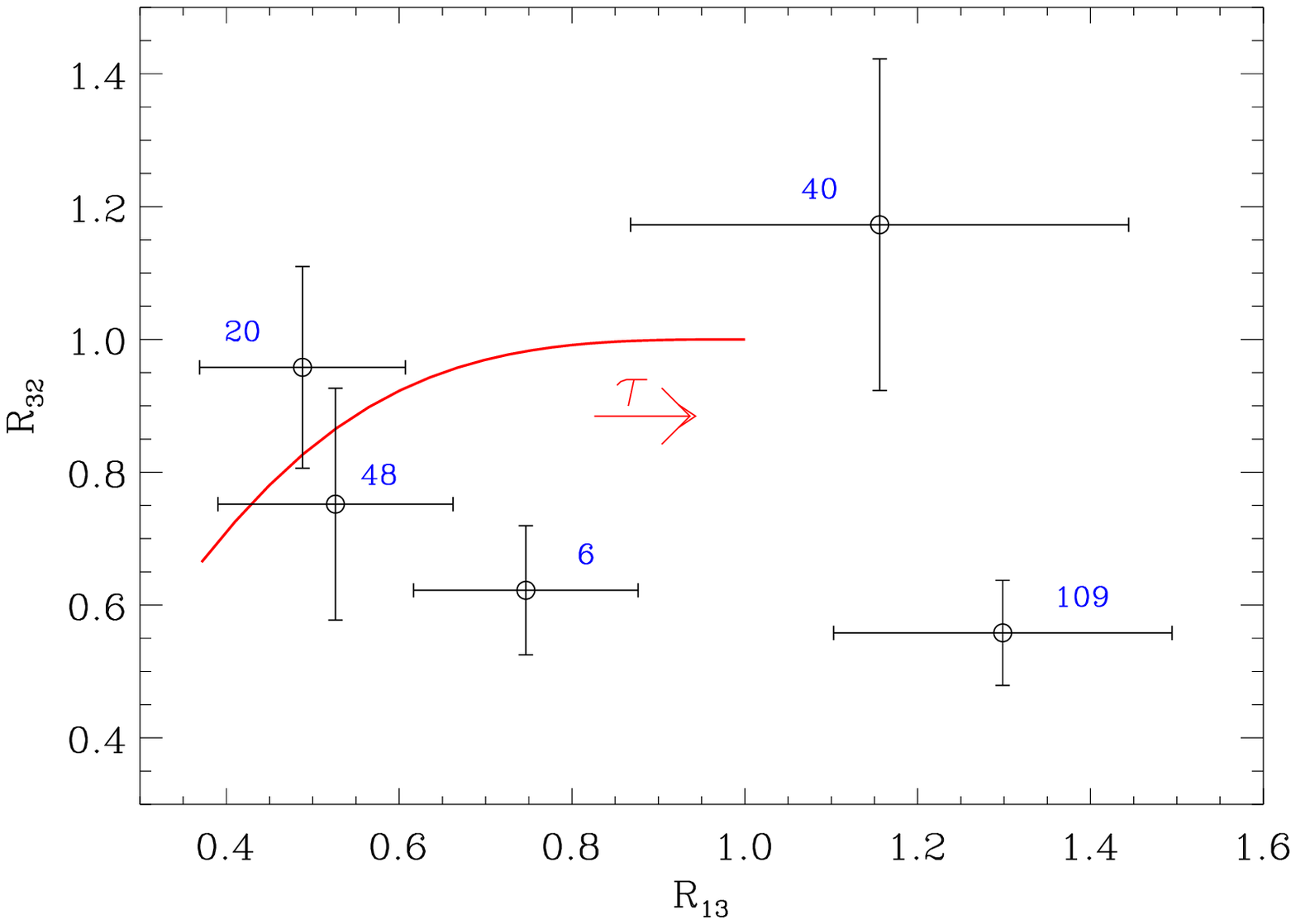}}

\caption{Ratio of the integrated intensities of couples of components of
\cdh~(1--0)  ({\it upper panel}, see Table~5 in \citealp{padovani09} for labels)
and HCN~(1--0) ({\it lower panel}, see Table~B.1 in \citealp{padovani11} for
labels). {\em Empty circles}:  observational data labeled with the respective core
number. {\em Red solid line}: one-slab LTE model, optical depth increases in the
arrow direction (see \citealp{padovani09} for details). \label{fig_LTE}}

\end{center}
\end{figure}

We followed the procedure developed in \citet{padovani09} to study the
departures from Local Thermodynamic Equilibrium (LTE) of two of the molecular
transitions with hyperfine components, \cdh~(1--0) and \hcn~(1--0), toward the
\pipe\ starless cores. By comparing ratios of integrated intensities between
couples of the $i$-th and $j$-th component, $R_{ij}$, it is possible to check
for opacity degree and LTE departures (Fig.~\ref{fig_LTE}). Under LTE and
optically thin conditions, the relative weightings of \cdh~(1--0) hyperfine 
components have the form 1:10:5:5:2:1, whereas for \hcn~(1--0) the relative
intensities are 3:5:1.   Figure~\ref{fig_LTE} suggests that cores~40 and 109
are the most optically thick, in agreement with the determination of $\tau$
from the HFS fit in CLASS (Table~\ref{tab_param1}), while the other cores are
optically thin. Core~20 is a particular case because it shows $R_{54}$  and
$R_{13}$ values in \cdh\ that cannot be reproduced with any value of $\tau$.
This can be  explained as the result of enhanced trapping due to an
overpopulation of the $(N,J,F)=(0,1/2,1)$ level, where most $N=1-0$ transitions
end (except for components 3 and 6; \citealp{padovani09}). This means that
these results  have to be thought in a qualitative way, since lines of very
different intrinsic intensities experience different balance between trapping
and collisions  leading to excitation anomalies. The hyperfine components of
\hcn~(1--0) do not obey the LTE weightings. For instance, as shown in
Fig.~\ref{fig:lineshfs}, core~6 is affected by strong auto-absorption of the
$F$=1--1 and $F$=2--1 components. Similarly, $F$=1--1 is stronger than $F$=2--1
in core~20. A more reliable determination of the \hcn\ abundance would be given
by the $^{13}$C (or $^{15}$N) isotopologue of \hcn\ \citep{padovani11}. In
general, cores seem to be close to LTE with those next to the optically thin
limit showing the smallest LTE departures.

\section{Discussion}\label{sec_discuss}

\subsection{Observational maps and physical structure of the
cores\label{dis_comp}}

The extinction maps show that the cores in the \pipe\ are surrounded by a 
diffuse medium (see Fig.~\ref{fig:pipe} and \citealp{lombardi06}).
\citet{roman11}  show that the 1.2 mm continuum MAMBO-II maps underestimate the
emission from the diffuse molecular component  due to the reduction algorithms
(see also \paper). To study this effect, we compared, at the center of the nine
cores, the $N_{\rm H_2}$ derived from the MAMBO-II maps
(Table~\ref{tab_dust_col_dens}) with the \Av\ value  from the extinction maps
of \citet{roman09, roman10}.  We found a statistically significant correlation
that can be expressed as 

\begin{equation}
\Av=(6.7\pm1.5)+(1.27\pm0.12)\times10^{-21} N_{\rm H_2}.
\label{eq_av}
\end{equation}

\noindent The proportionality factor is compatible with the standard value
(1.258$\times$10$^{-21}$~mag~cm$^{2}$; \citealp{wagenblast89}). However, the
comparison evidences that the 1.2~mm maps underestimate the column density in
average by an \Av\ of 6.7~mag, which  is likely the contribution from the
diffuse cloud material.  As a consequence, the \Av\ peak values of the cores
\citep[from][]{roman09, roman10} should be taken as upper limits of their
column densities.

The statistics of this study have increased with the whole nine core sample. In
Table~\ref{tab_resum} we show the main physical, chemical and polarimetric
properties of the starless cores with respect to core~109 ordered by its
decreasing \Av\ peak.  Column and volume density, and total mass tend to
decrease accordingly. On the contrary, the core diameter tends to increase. This
suggests that denser cores are smaller and more compact, which is
expected for structures in hydrostatic equilibrium such as Bonnor-Ebert spheres
(Frau et al., in prep.). Under such assumption, gravitationally
unstable cores (\nhd\gsim10$^5$~\cmt: \citealp{keto08}) would slowly condense
through a series of subsonic quasi-static equilibrium stages until the protostar
is born and gravitational collapse starts, while gravitationally stable cores
(\nhd\lsim10$^5$~\cmt) would achieve the equilibrium state and survive under
modest perturbations \citep{keto05}. The former group of cores would become
denser with time while developing an increasingly richer chemistry,
while the latter group would show a density-dependent chemistry (either in
terms of active chemical paths and excitation effects) stable in time. This
likely trends are supported by the clear correlation of the core chemistry
with the visual extinction peak of the core and, therefore, its central density
and structure (Section~\ref{dis_chem}). Regarding the increasing mass
with increasing central density, it seems unlikely that these young, diffuse
cores efficiently accrete mass from the environment. This trend is most likely
related to the initial conditions of formation of condensations in the low end
of the cloud mass spectrum: an initially more massive condensation is more
likely to form a dense structure.

   \begin{figure}[t]
   \centering
   \includegraphics[width=\columnwidth,angle=0]{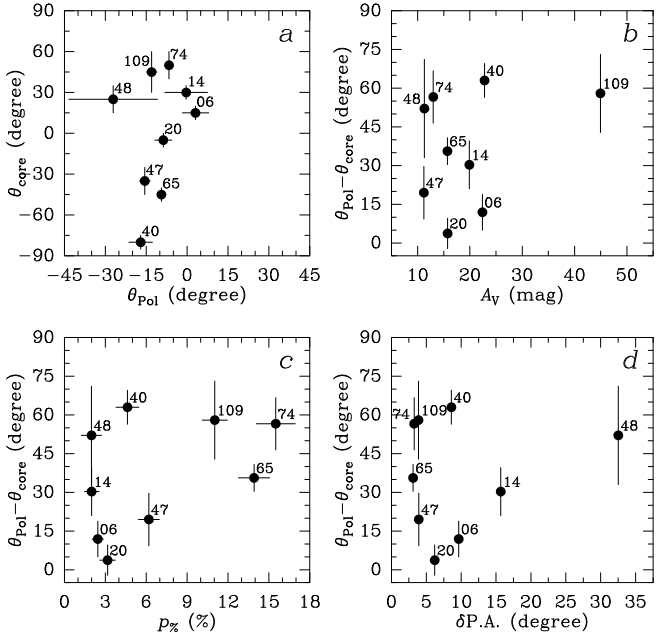}

\caption{Relation of the core major axis orientation ($\theta_{\rm core}$) to
different  physical parameters. {\it a}: $\theta_{\rm core}$ against mean
polarization angle ($\theta_{\rm Pol}$) of the corresponding observed field 
\citep{franco10}. {\it b-d}: $\theta_{\rm Pol}$-$\theta_{\rm core}$ as a
function of core \Av\ peak \citep{roman09,roman10}, polarization fraction
(\polp) and polarization angle dispersion (\deltapa), respectively.}

       \label{fig:orient}
   \end{figure}

\subsection{Relationship between the large scale magnetic field and the
elongation of the cores}

The \pipe\ cores are embedded in a sub-Alv\'enic molecular cloud that is
threaded with a strikingly uniform magnetic field \citep{alves08,franco10}.
Thus, it is possible that the core formation is related to the magnetic field
and its direction is related to the core elongations. Figure~\ref{fig:mambo} of
both \paper\ and the present work show that the polarization vectors calculated
from optical data cannot trace the densest part of the cores although the
vectors lie very close to the core boundaries (up to \Av$\sim$5~mag).
To derive the orientation of the core major axis, we computed the integrated
flux within the FWHM contour for a series of parallel strips (11\arcsec\ wide),
with position angles in the -90\deg\ to 90\deg\ range.  The major axis is
oriented in the direction with the largest integrated flux on the fewest strips.
We found no correlation between  the orientation of the major axis of the core,
$\theta_{\rm core}$, and the diffuse gas mean magnetic field
direction around the cores ($\theta_{\rm Pol}$, see Fig.~\ref{fig:orient}$a$).
To investigate more subtle effects, we compared for each core the difference
between polarization position angle and the major axis orientation ($\theta_{\rm
Pol}$-$\theta_{\rm core}$) with respect to the \Av\ peak, the polarization
fraction (\polp) and polarization angle dispersion (\deltapa) estimated in a
region of few arc-minutes around the cores \citep{franco10}. The results of
these comparisons are shown in panels $b$, $c$ and $d$ of Fig.~\ref{fig:orient},
respectively. Again, it seems that there are no clear correlations between these
quantities.

These results suggest that the well-ordered, large scale magnetic field that may
have driven  the gas to form the $\sim$15~pc long filamentary cloud, and shows
uniform properties for the diffuse gas at scales ranging \lsim0.01~pc up to several
pc \citep{franco10}, has little effect in shaping the morphology of the
denser $\sim$0.1~pc cores. At intermediate scales, \citet{peretto12} suggest
that a large scale compressive flow has contributed to the formation of a rich,
organized network of filamentary structures within the cloud, $\sim$0.1~pc wide and
up to a few pc long, which tend to align either parallel or perpendicular to the
magnetic field. If this is the case, then, rather than ambipolar diffusion, other
mechanisms such as a compressive flow should play a major role in the formation of
the \pipe\ cores. However, as pointed out by \citet{lada08}, the cores in the \pipe\
evolve on acoustic, and thus, slow timescales ($\sim$$10^6$~yr), allowing ambipolar
diffusion to have significant effects. Furthermore, the lack of spherical symmetry
demands an anisotropic active force. Projection effects, together with the small
statistical sample, require a deeper study of the magnetic field properties. In
order to extract firm conclusions and disentangle the nature of the
formation of the \pipe\ cores, the magnetic field toward the dense gas needs to be
studied.

\begin{table*}[t]
\caption{
\pipe\ Core General Properties with Respect to Core~109$^a$
}
\begin{tabular}{cccccccccccccc}\noalign{\smallskip}
\hline\hline\noalign{\smallskip}
\multicolumn{1}{c}{Source}& 
\multicolumn{1}{c}{Diameter}& 
\multicolumn{1}{c}{Mass}& 
\multicolumn{1}{c}{$N_{\rm H_2}$}& 
\multicolumn{1}{c}{$n_{\rm H_2}$}& 
\multicolumn{1}{c}{$p_\%^b$}& 
\multicolumn{1}{c}{$\delta$P.A.$^b$}& 
\multicolumn{1}{c}{X(C$^{18}$O)}& 
\multicolumn{1}{c}{X(CS)}& 
\multicolumn{1}{c}{X(C$_2$H)}& 
\multicolumn{1}{c}{X(C$_3$H$_2$)}& 
\multicolumn{1}{c}{X(CH$_3$OH)}& 
\multicolumn{1}{c}{X(CN)}& 
\multicolumn{1}{c}{X(N$_2$H$^+$)}\\
& \multicolumn{1}{c}{(pc)}& 
\multicolumn{1}{c}{($M_{\odot}$)}& 
\multicolumn{1}{c}{(10$^{21}$cm$^{-2}$)}& 
\multicolumn{1}{c}{(10$^{4}$cm$^{-3}$)}& 
\multicolumn{1}{c}{(\%)}& 
\multicolumn{1}{c}{($^o$)}& 
\multicolumn{1}{c}{(10$^{-11}$)}& 
\multicolumn{1}{c}{(10$^{-11}$)}& 
\multicolumn{1}{c}{(10$^{-11}$)}& 
\multicolumn{1}{c}{(10$^{-11}$)}& 
\multicolumn{1}{c}{(10$^{-11}$)}& 
\multicolumn{1}{c}{(10$^{-11}$)}& 
\multicolumn{1}{c}{(10$^{-11}$)}\\

Core 109 & 0.063 & 4.00 & 47.60 & 36.57 & 11.0 &  3.9 & 1324.6 &  40.2 &  34.1 &  52.8 &  44.3 &   8.7 &   2.2\\
\hline
\multicolumn{14}{c}{Relative Values (\%)} \\
Core 109 &  100 &  100 &  100 &  100 &  100 &  100 &  100 &  100 &  100 &  100 &  100 &  100 &  100\\
Core 40 &  165 &   63 &   23 &   14 &   42 &  222 &  517 &  168 &  335 &   52 &  272 &  480 &  208\\
\hline
Core 06 &   86 &   23 &   32 &   38 &   22 &  249 &  323 &  315 &  125 &   12 &  609 &  105 &  243\\
Core 14 &  113 &   35 &   28 &   25 &   18 &  404 & 1051 &  752 &   89 &    7 &  368 &  121 &   44\\
Core 20 &  141 &   28 &   14 &   10 &   29 &  160 &  752 &  576 &  295 & $<$   7 &  186 & $<$  36 & $<$  37\\
Core 65 &  105 &   26 &   23 &   22 &  126 &   80 &  608 & -- & $<$   3 & $<$   2 &   99 & $<$  13 & $<$  20\\
\hline
Core 74 &  154 &   16 &    7 &    4 &  140 &   84 &  303 &  435 & $<$   3 & $<$   5 &   96 & $<$  27 & $<$  36\\
Core 48 &  202 &   52 &   13 &    6 &   18 &  838 &  553 &  539 &  162 & $<$   2 &   72 & $<$  20 & $<$  25\\
Core 47 &  146 &   19 &    9 &    6 &   56 &  101 &  521 & -- &  298 &   13 &  204 &  511 &  117\\

\hline
\end{tabular}

\smallskip
$^a$ Molecules are ordered from earlier to later synthesization.
Cores are ordered in three groups following \paperd.\\
$^b$ \citet{franco10} \\
\label{tab_resum}

\end{table*}

   \begin{figure}[t]
   \centering
   \includegraphics[width=\columnwidth,angle=0]{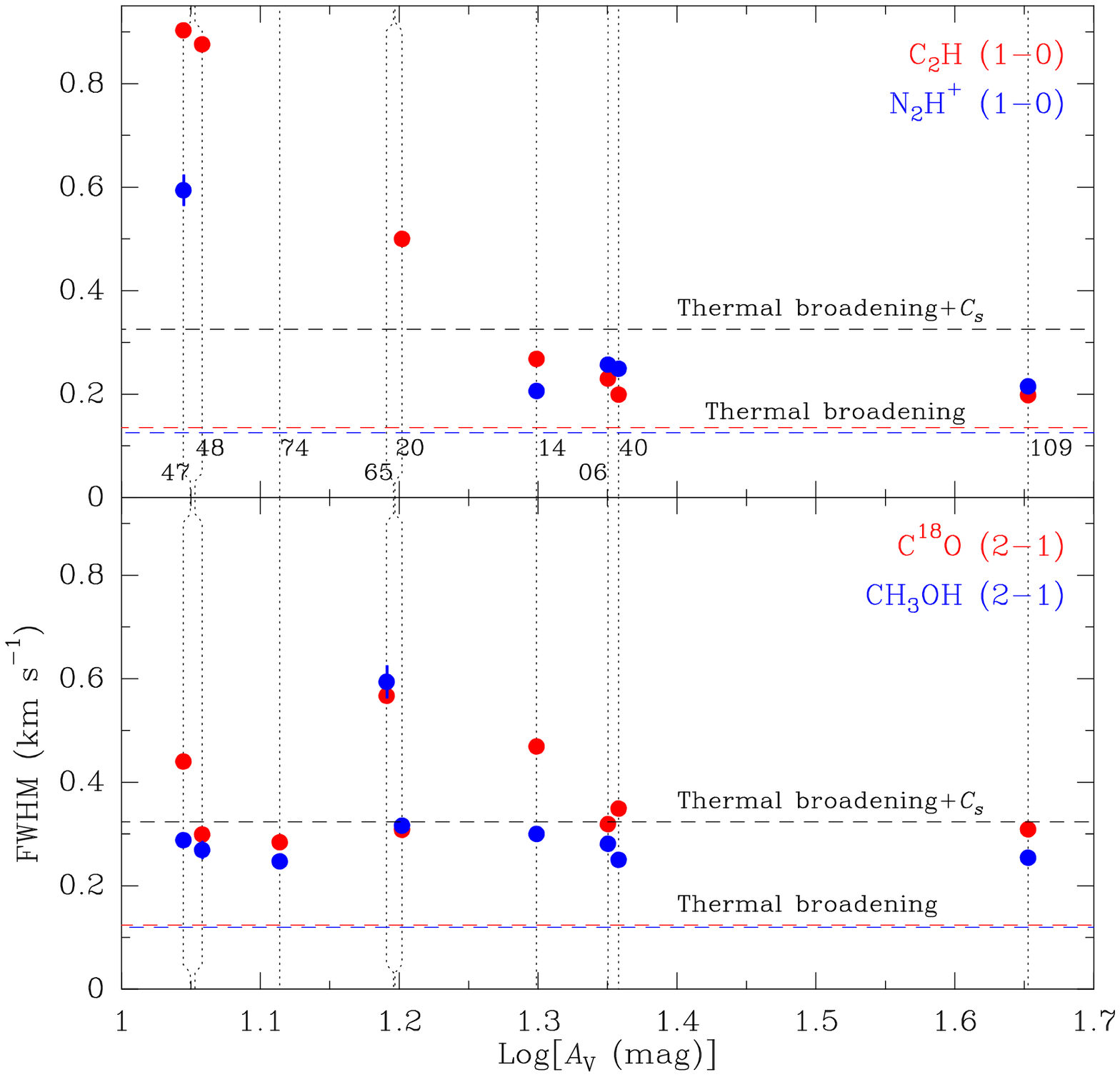}

      \caption{ \textit{Top panel}: line width of the \cdh~(1--0) and
\ndhp~(1--0) lines, colored in red and blue, respectively, as a function of the
logarithm of the core \Av\ peak. \textit{Bottom panel}: Same as top panel but for
the \cdo~(2--1) and \chtoh~(2--1) lines, colored in red and blue, respectively.
Colored dashed horizontal lines show the corresponding thermal line width for
each molecule for a temperature of 10 K. Black dashed horizontal lines show
the average thermal line width plus the sound speed ($C_s$) at 10~K. Vertical dotted lines mark
the \Av\ peak of each core, which in the case of similar values (cores 47 and 48,
and 20 and 65) have been slightly displaced. \label{fig:deltav}}

   \end{figure}

\subsection{Velocity dispersion analysis\label{DV}}

Figure~\ref{fig:deltav} shows the FWHM line width of four selected molecular 
transitions at 3~mm as a function of the core \Av\ peak.  Whereas the \cdo\ 
and \chtoh\ lines show an almost constant line width of 0.3--0.4~\kms\ for 
most cores (except core 65), the \cdh\ and \ndhp\ lines have a lower line
width, 0.20--0.25~\kms, except for the cores with lower visual  extinction (47,
48 and 20).  The values of the \cdo\ line are in agreement with  those found by
\citet{muench07} with lower angular resolution. In most cases,  the
line widths are only 2--3 times the thermal broadening at 10~K. These line
widths  imply a subsonic non-thermal velocity dispersion, $\sigma_{\rm NT}$, of
0.06--0.09~\kms\ for  the \ndhp\ and \cdh\ lines, respectively, and of
0.12--0.16 \kms\ for the \cdo\ and \chtoh\ lines, respectively.  Therefore, the
thermal pressure dominates the internal pressure  of the cores, which is a
general characteristic of the \pipe\ \citep{lada08}.  For the higher density
cores (\Av$>17$~mag), smaller $\sigma_{\rm NT}$ for the \ndhp\ and \cdh\ lines
with respect to \cdo\ and \chtoh\ lines suggests that the former transitions are
tracing the inner regions of the core. However, cores~47 and 48 present a
peculiar reverse case in the line width properties, i.e., the \cdh\ and \ndhp\
lines are significantly wider and clearly supersonic.  This is compatible with 
the plausible scenario of core~47 (and probably core 48) being a failed core 
in re-expansion \citep{frau12}, on which the centrally synthesized and 
initially narrow \ndhp\ and \cdh~(1--0) lines are now part of the disrupted
gas. But it is puzzling that the \cdo\ and \chtoh\ lines are still narrow and 
subsonic, unless they trace a part of core that still remains unperturbed. A
complete mapping of these cores is needed to reveal their striking nature.

\subsection{Discussion on the individual cores\label{ind-cores}}

Core~06, located in the western part of B59, is a compact and dense core. The
dust continuum emission is similar to the extinction maps \citep{roman09}.  The
core shows a rich chemistry with bright detections of all the early-time  and
some late-time molecules. The core has the brightest emission and highest 
abundance of \chtoh\ of the sample, as well as the highest \ndhp\ abundance. 
Unlike core~109, it has a high \cs\ abundance suggesting that it has not  been
depleted yet. All these features suggest that core~06 is in the 
``oxo-sulfurated'' group close to the ``deuterated'' cores.

Core~20, located in the {\it stem}, shows in the 1.2 mm map two components: a
compact and bright one surrounded by a second one, extended and diffuse. Most of
the early-time molecules were detected, and thus, this core seems to be very
young chemically showing abundances in \cs\ and \chtoh\ among the highest.
Normalized intensities are in general quite large for its density
(Fig.~\ref{fig_norm}), and it has a very large \so\ normalized integrated
intensity as core~47 (Fig.~\ref{fig_areaNorm}). These signposts suggest that
core~20 belongs to the ``oxo-sulfurated'' group.

Core~47, located between the \stem\ and the \bowl, has  extended and diffuse dust
emission. It shows a fairly uniform and weak emission all over the \Av\ and
MAMBO-II maps. This can explain the peak position difference of $\sim$130\arcsec\ 
between the dust emission map and the position taken by \citet{muench07} for the
line observations. It shows weak line emission, only in early-time molecules.  It
is the second least dense core of the sample ($\sim$2$\times$10$^{4}$~\cmt), yet
the molecular abundances tend to be among the highest. \citet{rathborne08}  report
a clear detection of \nht~(1,1) and hints of emission in the (2,2) transition. 
Figure~\ref{fig:lineshfs} shows a marginal detection at 3$\sigma$ level of the
\ndhp~(1--0) brightest hyperfine component. The high  molecular abundances and the
emission of certain molecular line tracers of the ``oxo-sulfurated'' group,
together with its diffuse morphology and low density typical of the ``diffuse''
cores, suggest that core~47 may be an evolved failed core now in re-expansion as
already suggested in \paperd. The relatively broad lines in some of the species
(see Section~\ref{DV}) support this scenario.

Core~65, located in the \bowl, is the central core of a  group of three (see
Fig.~\ref{fig:mambo}).  Its density is in the limit between those of the 
``diffuse'' and ``oxo-sulfurated'' cores.  It has a very poor chemistry with
only  ``ubiquitous''  early-time molecules detected (\co, \cs, and \chtoh) with
abundances among the lowest of the sample. The line widths, $\sim$0.6~\kms,
appear to be larger than those of the other cores.

Finally, core~74, located in the \bowl, is extended and diffuse similarly to
core~47. It also shows a very poor chemistry with only ``ubiquitous''
early-time molecules detected (\co, \cs, and \chtoh).

It is useful to review the data from \citet{rathborne08}. The late-time
molecule \nht\ in the (1,1) transition is detected in cores 06, 20, 47, and
marginally in 65. These cores belong to the ``oxo-sulfurated'' group, which
suggests that \nht\ is formed in this phase. \ccs\ is considered an early-time
molecule with a lifetime of \lsim \tenpow{3}{4}~yr \citep{degregorio06}. It is
only marginally detected toward core~06, therefore suggesting that it might be
very young. Core~74 does not show any emission, in agreement with the poor
chemistry detected in our 3~mm surveys. These results also suggest that the
five cores are in a very early stage of evolution.

\subsection{Qualitative chemistry discussion}
\label{dis_chem}

The molecular transitions from \paper\ and this work increase the number of
typical lines of the core categories established in \paperd. Four of the five
cores have lower densities than the initial subsample (except for core 48, 
Table~\ref{tab_dust_col_dens}), and thus, we are now including in the analysis
shallower cores that might be more affected by the external radiation field and
that show a younger chemistry  (although the timescale to form the core may
influence this evolution: \citealp{tafalla04,crapsi05}).  

We found a complex chemical scenario toward the \pipe\ cores. However, as
pointed out in \paperd, it seems that there is a chemical trend with
density in the form of three differentiated core chemical groups.  We remind
that the \Av\ values should be interpreted as upper limits for the column
density of the cores (Section~\ref{dis_comp}), and that the column densities
derived from the dust emission maps show larger differences  than the \Av\
values. These facts translate to larger abundance differences among the cores as
compared to the normalized intensity differences. The molecular trends, however,
are compatible. We will base our analysis in the combination of the results
obtained via the normalized intensities and normalized integrated intensities
(Figs.~\ref{fig_norm} and \ref{fig_areaNorm}), and of the molecular abundances
with respect to the H$_2$ (Table~\ref{tab_resum}), to which we will refer
generically as \mis. \cdo\ and \cs\ lines appear to be  ``ubiquitous'', as they
are detected in all the cores. Their \mis\ decrease with column density due to,
probably, an increasingly efficient depletion for both species (and
isotopologues) as density grows.  The variation of the \cn\ \mi\ among the cores
has increased with respect to \paper\ (up to a factor of $\sim$33 in abundance:
Table~\ref{tab_abun2}), due to the addition of more diffuse cores. The lower
limits are indeed very low, and thus, we are now exploring even younger chemical
stages of these starless cores. All these features suggest that \cn~(1--0) is a
transition typical of the ``deuterated'' group.  A nitrogen- (\ndhp) and a
deuterium-bearing (\cthdt) species, and a carbon chain molecule (\hctn) are 
shown in Fig.~\ref{fig_areaNorm}. These late-time molecules, present toward the
densest objects, are not detected in low density cores. They are only present
after achieving a density threshold, and exhibit increasing \mis\ as density
grows. These transitions seem to be typical of the ``deuterated'' cores, which
is consistent with the detection  of \nhdd~(1$_{1,1}$--1$_{0,1}$) toward this
core group in \paperd.

In order to analyze possible excitation effects in the detected emission
we consider now the \ndhp~(1--0) transition whose critical density ($n_{\rm
cr}$=\tenpow{2}{5}~\cmt: \citealp{ungerechts97}) lies within the density range of
the studied cores. The different excitation conditions could explain the differences
between the densest core (109) and the least dense cores with no detections (20, 48,
and 74), having the latter ones \nhd\ (toward the central beam derived from
Table~\ref{tab_dust_col_dens}) close to $n_{\rm cr}$. Indeed, we used RADEX
\citep{vandertak07} assuming as representative values $X\left({\rm
N_2H^+}\right)$=\tenpow{2}{-11}, $T$=10~K, and $\sigma$=0.25~\kms, and found that a
volume density increase from \tenpow{4}{5}~\cmt\ to \tenpow{1.4}{6}~\cmt\ (average
values toward the central beam of cores~6 and 109, respectively) produces a
difference in \Tmb\ of a factor of 3 while the observed peaks are one order of
magnitude apart, suggesting real differences in the abundances rather than
excitation effects. The observational classification proposed, although based on
groups of molecules and peak \Av\ values, is sensitive to these effects as the \Av\
values are related to those of \nhd. However, a more careful study should be done
when studying individual molecules to be compared to chemical modelling results.

\chtoh\ deserves a special mention.  This molecule is clearly detected in the
gas phase toward all the observed cores in the (2$_{0,2}$--1$_{0,1}$) (shown
normalized in Fig.~\ref{fig_norm}) and (2$_{1,2}$--1$_{1,1}$) transitions
(Fig.~\ref{fig:lineshfs_new}). It shows a behavior similar to that of the
``oxo-sulfurated''  species but peaking at slightly larger densities. Thus,
this species is likely to peak in the transition from the ``oxo-sulfurated''
core chemistry to the typical dense core chemistry found toward the
``deuterated'' cores, suggesting that \chtoh\ could be actually an early-time
molecule. It is expected to be formed efficiently in grain surfaces, with
abundances for the gas phase of $\sim$10$^{-9}$ at most
\citep{cuppen09,garrod11}, very close to the observational abundances derived
(\tenpow{$\sim$3}{-10}--\tenpow{$\sim$3}{-9}: Table~\ref{tab_abun1}). 
Abundances for the gas phase of \tenpow{6}{-10}, comparable to the lowest
values for the \pipe\ cores, have been derived in the literature through
modeling of more evolved low-mass cores \citep{tafalla06}. However, the higher
densities and comparable temperatures product of this modeling with respect to
the \pipe\ core values suggest that other mechanisms are needed to explain the
high gas phase \chtoh\ abundances found here. In addition, the abundances in
the \pipe\ cores seem to correlate with their location in the cloud, being
larger in the B59 region and decreasing as going toward the \bowl. This fact
could be explained by the slightly higher temperatures reported toward the B59
region \citep{rathborne08}, which could enhance evaporation from grains.

In summary, our high spectral resolution dataset shows the existence of a clear
chemical differentiation toward the \pipe\ cores. The chemical signatures agree
with the results of previous \paperud. Chemistry seems to become more rich and
complex as cores grow denser therefore suggesting an evolutionary gradient
among the sample. The tentative correlation found in \paper\ between magnetic
field and chemical evolutionary stage of the cores is less clear with the whole
nine core sample.


\section{Summary and conclusions}

We carried out observations of continuum and line emission toward five starless
cores, located on the three different regions of the \pipe, and combined them
with the observations of the four additional cores of \paper\ to extend the
dataset to nine cores. We studied the physical and chemical properties of the
cores, and their correlation following \paperd. We also studied the correlation
with the magnetic field properties of the surrounding diffuse gas following
\paper.

\begin{enumerate}

\item The \pipe\ starless cores show very different morphologies.  The complete
sample of nine cores contains dense and compact cores (6, 65, and 109;
\nhd\gsim10$^5$~\cmt), diffuse and elliptical/irregular ones (20, 40, 47, 48,
and 74; \nhd\lsim5$\times$10$^4$~\cmt), and a filament containing the relatively
dense core~14 (\nhd$\sim$9$\times$10$^4$~\cmt). The average properties of the
nine cores of the sample are diameter of $\sim$0.08~pc ($\sim$16,800~AU),
density of $\sim$10$^5$~\cmt, and mass of $\sim$1.7~\msun. These values are very
close to (but less dense than) those reported by
\citet{wardthompson99} for a set of very young dense cores and, therefore,
typical of even earlier stages of evolution.

\item MAMBO-II maps are in a general good morphological agreement with previous
extinction maps \citep{lombardi06}. By comparing the \Av\ peak values of the
nine cores from deeper NICER maps \citep{roman09,roman10}, we derived a
proportionality factor \Av/\Nhd=(1.27$\pm$0.12)$\times$10$^{-21}$~mag~cm$^{2}$,
compatible with the standard value (\tenpow{1.258}{-21}~mag~cm$^{2}$;
\citealp{wagenblast89}). In addition, we found that dust continuum maps
underestimate the column density by an \Av\ of $\sim$6.7~mag that may be
arising from the diffuse material of the cloud.

\item The orientation of the cores is not correlated with the surrounding
diffuse gas magnetic field direction, which suggests that large scale
magnetic fields are not important in shaping the cores. On the other hand, the lack
of spherical symmetry demands an important anisotropic force, and
projection effects might be important. A deeper study of the magnetic field
of the dense gas is needed.

\item The analysis of the line widths reports two behaviors depending on the
molecular transition: ({\it i}) a roughly constant value of subsonic turbulent
broadening for all the cores (e.g. \cdo~(1--0) and \chtoh~(2--1), see also
\citealp{lada08}) and ({\it ii}) a roughly constant slightly narrower
broadening for cores with \Av\gsim20~mag and supersonic turbulent broadenings
otherwise (e.g. \cdh~(1--0) and \ndhp~(1--0)).

\item We observed a set of early- and late-time molecular transitions toward
the cores and derived their column densities and abundances.  The high spectral
resolution molecular normalized line data is in agreement  with the lower
spectral resolution data presented in \paperd.  The nine starless cores are all
very chemically young but show different chemical properties. ``Diffuse'' cores
(\Av\lsim15~mag: 48 and 74) show emission only in ``ubiquitous'' lines typical
of the parental cloud chemistry (e.g. \co, \cs, \chtoh). The denser
``deuterated'' cores (\Av\gsim22~mag: 40 and 109) show weaker \mis\ for
``ubiquitous'' lines and present emission in nitrogen- (\ndhp) and
deuterium-bearing (\cthdt) molecules, and in some carbon chain molecules
(\hctn), signposts of a prototypical dense core chemistry. ``Oxo-sulfurated''
cores (\Av$\simeq$15--22~mag: 6, 14, 20, and 65) are in a chemical transitional
stage between cloud and dense core chemistry. They are characterized by
presenting large \mis\ of \chtoh\ and oxo-sulfurated molecules (e.g. \so\ and
\sod) that disappear at higher densities, and they still present significant
emission in the ``ubiquitous'' lines. \chtoh\ was detected toward the nine
cores of the complete sample with abundances of $\sim$10$^{-9}$, close to the
maximum value expected for gas-phase chemistry.

\item Core 47 presents high abundance of \chtoh\ and \ndhp, in spite of being
the core with the lowest \hd\ column density, and broad line width in some
species  (\cdh\ and \ndhp). All this is in agreement with the hypothesis given
in \paperd, which suggests that Core~47 could be a failed core. 

\item The chemical evolutionary stage is not correlated with the core location
in the \pipe, but it is correlated with the physical properties of the cores
(density and size). Thus, the chemically richer cores are the denser ones.  The
tentative correlation between magnetic field and chemical properties found for
the initial subsample of four cores is  less clear with the current sample.  

\end{enumerate}

The \pipe\ is confirmed as an excellent laboratory for studying the very early
stages of star formation. The nine cores studied show different morphologies
and different chemical and magnetic properties. Physical and chemical
properties seem to be related, although important differences arise, which
evidence the complex interplay among thermal, magnetic, and turbulent energies
at core scales. Therefore, a larger statistics is needed to better understand
and characterize the \pipe\ starless core evolution. In addition, other young
clouds with low-mass dense cores, such as the more evolved star-forming Taurus
cloud, should be studied in a similar way to prove the presented results as a
general trend or, on the contrary, a particular case for a filamentary
magnetized cloud.


\begin{acknowledgements}

The authors want to acknowledge all the IRAM 30-m staff for their
hospitality during the observing runs, the operators and AoDs for their active
and invaluable support, G.~Quintana-Lacaci for his help during the observing and
reduction process of MAMBO-II data, J. Kauffmann for kindly helping on the
implementation of his MAMBO-II new reduction scheme, and C. Rom\'an-Z\'u\~niga
for gently allowing us to make use of the NICER maps.
P.F. was partially supported by MINECO predoctoral fellowship FPU (Spain).
P.F., J.M.G., M.P., F.O.A., and R.E. are partially supported by MINECO grants
AYA2008-06189-C03-02 and AYA2011-30228-C03-02 (Spain), as well as by AGAUR 
grant 2009SGR1172 (Catalonia).
F.O.A. is also supported by Deutsche Forschungsgemeinschaft (DFG) through the
Emmy Noether Research grant VL 61/3-1 and through SFB 956.
G.B. is funded by an Italian Space Agency (ASI) fellowship under contract number
I/005/07/01. G.A.P.F. is partially supported by CNPq and FAPEMIG (Brazil). O.M.
is supported by the NSC (Taiwan) ALMA-T grant to the Institute of Astronomy \&
Astrophysics, Academia Sinica. 
We made extensive use of NASA's Astrophysics Data System (NASA/ADS).

\end{acknowledgements}


\appendix

\section{On line material: Tables and Figures}

\begin{table*}
\caption{
Line parameters$^a$.
}
\begin{tabular}{cc r@{.}l r@{.}l r@{.}l r@{.}l r@{.}l r@{.}l c }
\noalign{\smallskip}
\hline\hline\noalign{\smallskip}
\multicolumn{1}{c}{Molecular} &
& \multicolumn{2}{c}{$T_{\rm MB}$ $^b$} 
& \multicolumn{2}{c}{$A \times \tau$ $^c$} 
& \multicolumn{2}{c}{$\int T_{\rm MB}\mbox{d}\texttt{v}$ $^b$} 
& \multicolumn{2}{c}{$\texttt{v}_{\rm LSR}$} 
& \multicolumn{2}{c}{$\Delta\texttt{v}_{\rm LSR}$} 
\\
\multicolumn{1}{c}{transition}
& \multicolumn{1}{c}{Source} 
& \multicolumn{2}{c}{(K)} 
& \multicolumn{2}{c}{(K)} 
& \multicolumn{2}{c}{(K\, km\,s$^{-1}$)} 
& \multicolumn{2}{c}{(km\,s$^{-1}$)} 
& \multicolumn{2}{c}{(km\,s$^{-1}$)}
& \multicolumn{2}{c}{$\tau$ $^d$}
& \multicolumn{1}{c}{Profile$^e$} \\
\hline
C$_3$H$_2$~(2$_{1,2}$--1$_{1,0}$)
 & Core 06 &    0&502(24) & \multicolumn{2}{c}{--} &    0&140(5) &    3&574(4) &    0&262(11) & \multicolumn{2}{c}{--} & G\\
 & Core 14 &    0&37(6) & \multicolumn{2}{c}{--} &    0&086(11) &    3&502(14) &    0&22(3) & \multicolumn{2}{c}{--} & G\\
 & Core 40 &    1&19(5) & \multicolumn{2}{c}{--} &    0&347(9) &    3&420(4) &    0&273(9) & \multicolumn{2}{c}{--} & G\\
 & Core 47 &    0&079(22) & \multicolumn{2}{c}{--} &    0&071(8) &    3&11(5) &    0&85(8) & \multicolumn{2}{c}{--} & G\\
 & Core 109 &    2&74(6) & \multicolumn{2}{c}{--} &    0&799(13) &    5&8340(20) &    0&274(5) & \multicolumn{2}{c}{--} & G\\
\hline
C$_2$H~(1--0)
 & Core 06 & \multicolumn{2}{c}{--} &    0&389(7) & \multicolumn{2}{c}{--} &    3&6200(8) &    0&2300(19) &    0&655(21) & G\\
 & Core 14 & \multicolumn{2}{c}{--} &    0&1650(24) & \multicolumn{2}{c}{--} &    3&5800(13) &    0&268(3) &    0&450(9) & G\\
 & Core 20 & \multicolumn{2}{c}{--} &    0&1140(25) & \multicolumn{2}{c}{--} &    3&7900(19) &    0&500(6) &    0&193(16) & G\\
 & Core 40 & \multicolumn{2}{c}{--} &    2&03(3) & \multicolumn{2}{c}{--} &    3&4700(4) &    0&1990(8) &    2&58(4) & G\\
 & Core 47 & \multicolumn{2}{c}{--} &    0&0345(5) & \multicolumn{2}{c}{--} &    3&140(7) &    0&903(15) &    0&1000(4) & G\\
 & Core 48 & \multicolumn{2}{c}{--} &    0&0255(5) & \multicolumn{2}{c}{--} &    3&630(9) &    0&876(21) &    0&1000(16) & G\\
 & Core 109 & \multicolumn{2}{c}{--} &    2&280(9) & \multicolumn{2}{c}{--} &    5&89000(13) &    0&1980(3) &    1&530(8) & G\\
\hline
CH$_{3}$OH~(2$_{0,2}$--1$_{0,1}$)
 & Core 06 &    1&841(15) & \multicolumn{2}{c}{--} &    0&55(3) &    3&512(7) &    0&281(15) & \multicolumn{2}{c}{--} & G\\
 & Core 14 &    1&30(3) & \multicolumn{2}{c}{--} &    0&416(21) &    3&519(7) &    0&300(17) & \multicolumn{2}{c}{--} & G\\
 & Core 20 &    0&43(3) & \multicolumn{2}{c}{--} &    0&145(10) &    3&653(10) &    0&316(24) & \multicolumn{2}{c}{--} & G\\
 & Core 40 &    1&230(15) & \multicolumn{2}{c}{--} &    0&327(16) &    3&375(6) &    0&250(14) & \multicolumn{2}{c}{--} & G\\
 & Core 47 &    0&45(3) & \multicolumn{2}{c}{--} &    0&138(10) &    2&845(10) &    0&288(24) & \multicolumn{2}{c}{--} & G\\
 & Core 48 &    0&199(25) & \multicolumn{2}{c}{--} &    0&057(6) &    3&652(13) &    0&27(3) & \multicolumn{2}{c}{--} & G\\
 & Core 65 &    0&15(3) & \multicolumn{2}{c}{--} &    0&098(9) &    5&04(3) &    0&59(6) & \multicolumn{2}{c}{--} & G\\
 & Core 74 &    0&257(21) & \multicolumn{2}{c}{--} &    0&068(5) &    4&201(9) &    0&247(22) & \multicolumn{2}{c}{--} & G\\
 & Core 109 &    1&27(3) & \multicolumn{2}{c}{--} &    0&343(18) &    5&778(6) &    0&254(15) & \multicolumn{2}{c}{--} & G\\
\hline
CH$_{3}$OH~(2$_{-1,2}$--1$_{-1,1}$)
 & Core 06 &    1&432(15) & \multicolumn{2}{c}{--} &    0&417(4) &   3&5055(10) &    0&273(3) & \multicolumn{2}{c}{--} & G\\
 & Core 14 &    1&04(3) & \multicolumn{2}{c}{--} &    0&306(6) &    3&508(3) &    0&276(7) & \multicolumn{2}{c}{--} & G\\
 & Core 20 &    0&35(3) & \multicolumn{2}{c}{--} &    0&094(6) &    3&672(8) &    0&250(20) & \multicolumn{2}{c}{--} & G\\
 & Core 40 &    0&998(15) & \multicolumn{2}{c}{--} &    0&252(3) &   3&3705(10) &    0&237(3) & \multicolumn{2}{c}{--} & G\\
 & Core 47 &    0&33(3) & \multicolumn{2}{c}{--} &    0&119(7) &    2&847(10) &    0&344(22) & \multicolumn{2}{c}{--} & G\\
 & Core 48 &    0&134(25) & \multicolumn{2}{c}{--} &    0&047(5) &   3&641(19) &    0&33(5) & \multicolumn{2}{c}{--} & G\\
 & Core 65 &    0&12(3) & \multicolumn{2}{c}{--} &    0&063(7) &   4&98(3) &    0&48(6) & \multicolumn{2}{c}{--} & G\\
 & Core 74 &    0&212(21) & \multicolumn{2}{c}{--} &    0&059(4) &   4&201(9) &    0&261(20) & \multicolumn{2}{c}{--} & G\\
 & Core 109 &   1&03(3) & \multicolumn{2}{c}{--} &    0&263(5) &   5&7705(20) &    0&240(6) & \multicolumn{2}{c}{--} & G\\
\hline
CN~(1--0)
 & Core 06 & \multicolumn{2}{c}{--} &    0&17(5) & \multicolumn{2}{c}{--} &    3&640(15) &    0&30(3) &    1&2(4) & G\\
 & Core 14 & \multicolumn{2}{c}{--} &    0&051(9) & \multicolumn{2}{c}{--} &    3&64(8) &    0&81(15) &    0&1(7) & G\\
 & Core 40 & \multicolumn{2}{c}{--} &    0&65(22) & \multicolumn{2}{c}{--} &    3&430(21) &    0&36(5) &    3&9(1.3) & G\\
 & Core 47 & \multicolumn{2}{c}{--} &    0&10(4) & \multicolumn{2}{c}{--} &    2&98(5) &    0&80(13) &    0&9(5) & G\\
 & Core 109 & \multicolumn{2}{c}{--} &    1&41(22) & \multicolumn{2}{c}{--} &    5&930(5) &    0&162(11) &    1&13(23) & G\\
 &  & \multicolumn{2}{c}{--} &    2&3(1.3) & \multicolumn{2}{c}{--} &    5&670(7) &    0&101(16) &    4&(3) & \\
\hline
HCN~(1--0)
 & Core 06 & \multicolumn{2}{c}{--} &    0&025(6) & \multicolumn{2}{c}{--} &    3&56(3) &    0&76(8) &    0&11(5) & G\\
 & Core 20 & \multicolumn{2}{c}{--} &    0&059(16) & \multicolumn{2}{c}{--} &    3&58(3) &    0&68(7) &    0&24(8) & G\\
 & Core 40 & \multicolumn{2}{c}{--} &    1&55(11) & \multicolumn{2}{c}{--} &    3&410(16) &    0&334(22) &    6&0(5) & NS\\
 & Core 47 & \multicolumn{2}{c}{--} &    0&051(13) & \multicolumn{2}{c}{--} &    2&93(3) &    0&72(7) &    0&27(8) & G\\
 & Core 48 & \multicolumn{2}{c}{--} &    0&33(10) & \multicolumn{2}{c}{--} &    3&54(5) &    0&90(11) &    2&4(1.2) & G\\
 & Core 109 & \multicolumn{2}{c}{--} &    2&53(3) & \multicolumn{2}{c}{--} &    5&93(7) &    0&16(22) &    0&25(10) & NS\\
 &  & \multicolumn{2}{c}{--} &    6&10(3) & \multicolumn{2}{c}{--} &    5&72(7) &    0&25(22) &   10&20(10) & \\
\hline
N$_2$H$^+$~(1--0)
 & Core 06 & \multicolumn{2}{c}{--} &    0&119(5) & \multicolumn{2}{c}{--} &   3&5000(16) &    0&257(4) &    0&10(10) & G\\
 & Core 14 & \multicolumn{2}{c}{--} &    0&0341(16) & \multicolumn{2}{c}{--} &   3&500(5) &    0&206(10) &    0&10(9) & G\\
 & Core 40 & \multicolumn{2}{c}{--} &    0&219(12) & \multicolumn{2}{c}{--} &   3&4000(19) &    0&249(5) &    0&171(25) & G\\
 & Core 47 & \multicolumn{2}{c}{--} &    0&0100(9) & \multicolumn{2}{c}{--} &   3&00(4) &    0&59(6) &    0&10(3) & G\\
 & Core 109 & \multicolumn{2}{c}{--} &    0&904(14) & \multicolumn{2}{c}{--} &   5&8000(5) &    0&2150(11) &    0&467(11) & G\\
\hline
N$_2$D$^+$~(2--1)$^f$
 & Core 40 &    0&084(20) & \multicolumn{2}{c}{--} &    0&019(3) &    3&280(15) &    0&21(3) & \multicolumn{2}{c}{--} & G\\
 & Core 109 &    0&31(4) & \multicolumn{2}{c}{--} &    0&109(7) &    5&673(11) &    0&331(22) & \multicolumn{2}{c}{--} & G\\
\hline
DCO$^+$~(3--2)
 & Core 06 &    0&44(13) & \multicolumn{2}{c}{--} &    0&22(3) &    3&58(3) &    0&48(11) & \multicolumn{2}{c}{--} & G\\
 & Core 109 &    0&70(11) & \multicolumn{2}{c}{--} &    0&151(18) &    5&828(13) &    0&202(21) & \multicolumn{2}{c}{--} & G\\
\hline
\hline
\end{tabular}
\smallskip

$^a$ Line parameters of the detected lines. Multiple velocity components are
shown if present. For the molecular transitions with no 
hyperfine components, the parameters for the transitions labeled as G (last
column) have been derived from a Gaussian fit while line parameters of NS and SA
profiles have been derived from the intensity peak ($T_{\rm MB}$), and zero
(integrated intensity), first (line velocity) and second (line width) order moments
of the emission.  For the molecular transitions with hyperfine components, the
parameters have been derived using the hyperfine component fitting method of the
CLASS package. The value in parenthesis shows the uncertainty of the last digit/s. If
the two first significative digits of the error are smaller than 25, two digits are
given to better constrain it.\\
$^b$ Only for molecular transitions with no hyperfine components.\\
$^c$ Only for molecular transitions with hyperfine components.\\
$^d$ Derived from a CLASS hyperfine fit for molecular transitions with hyperfine 
components.
Derived numerically for CS, C$^{34}$S, $^{13}$CO, and C$^{18}$O using 
Eq.~1 from \paper. A value of 0.1 is assumed when
no measurement is available.\\
$^e$ G: Gaussian profile. NS: Non-symmetric profile. SA: Self-absorption profile.\\
$^f$ Only the main component is detected.\\
\label{tab_param1}
\end{table*}

\begin{table*}
\caption{
Line parameters$^a$ (Continuation).
}
\begin{tabular}{cc r@{.}l r@{.}l r@{.}l r@{.}l r@{.}l r@{.}l c }
\noalign{\smallskip}
\hline\hline\noalign{\smallskip}
\multicolumn{1}{c}{Molecular} &
& \multicolumn{2}{c}{$T_{\rm MB}$ $^b$} 
& \multicolumn{2}{c}{$A \times \tau$ $^c$} 
& \multicolumn{2}{c}{$\int T_{\rm MB}\mbox{d}\texttt{v}$ $^b$} 
& \multicolumn{2}{c}{$\texttt{v}_{\rm LSR}$} 
& \multicolumn{2}{c}{$\Delta\texttt{v}_{\rm LSR}$} 
\\
\multicolumn{1}{c}{transition}
& \multicolumn{1}{c}{Source} 
& \multicolumn{2}{c}{(K)} 
& \multicolumn{2}{c}{(K)} 
& \multicolumn{2}{c}{(K\, km\,s$^{-1}$)} 
& \multicolumn{2}{c}{(km\,s$^{-1}$)} 
& \multicolumn{2}{c}{(km\,s$^{-1}$)}
& \multicolumn{2}{c}{$\tau$ $^d$}
& \multicolumn{1}{c}{Profile$^e$} \\
\hline
C$^{34}$S~(2--1)
 & Core 06 &    0&207(16) & \multicolumn{2}{c}{--} &    0&055(3) &    3&551(7) &    0&247(15) &    0&182(18) & G\\
 & Core 14 &    0&267(25) & \multicolumn{2}{c}{--} &    0&068(5) &    3&545(8) &    0&241(20) &    0&48(5) & G\\
 & Core 20 &    0&18(4) & \multicolumn{2}{c}{--} &    0&057(7) &    3&718(19) &    0&30(4) &    0&154(15) & G\\
 & Core 40 &    0&268(16) & \multicolumn{2}{c}{--} &    0&069(3) &    3&381(5) &    0&241(13) &    0&140(14) & G\\
 & Core 47 &    0&07(4) & \multicolumn{2}{c}{--} &    0&042(10) &    3&00(7) &    0&53(13) &    0&036(4) & G\\
 & Core 48 &    0&187(23) & \multicolumn{2}{c}{--} &    0&041(4) &    3&729(11) &    0&20(3) &    0&26(3) & G\\
 & Core 74 &    0&137(21) & \multicolumn{2}{c}{--} &    0&027(3) &    4&236(12) &    0&186(23) &    0&228(23) & G\\
 & Core 109 &    0&34(3) & \multicolumn{2}{c}{--} &    0&083(5) &    5&825(7) &    0&233(17) &    0&185(18) & G\\
\hline
CS~(2--1)
 & Core 06 &    1&20(6) & \multicolumn{2}{c}{--} &    0&30(3) &    3&429(11) &    0&237(21) &    4&1(4) & G\\
 &  &    0&56(6) & \multicolumn{2}{c}{--} &    0&14(3) &    3&698(22) &    0&23(5) & \multicolumn{2}{c}{--} & \\
 & Core 14 &    0&69(10) & \multicolumn{2}{c}{--} &    0&41(3) &    3&439(21) &    0&45(4) &   10&7(1.1) & SA\\
 & Core 20 &    1&18(9) & \multicolumn{2}{c}{--} &    0&34(4) &    3&469(12) &    0&27(3) &    3&5(3) & G\\
 &  &    1&10(9) & \multicolumn{2}{c}{--} &    0&35(4) &    3&820(14) &    0&30(3) & \multicolumn{2}{c}{--} & \\
 & Core 40 &    1&94(7) & \multicolumn{2}{c}{--} &    0&560(17) &    3&369(4) &    0&415(14) &    3&1(3) & NS\\
 & Core 47 &    1&16(9) & \multicolumn{2}{c}{--} &    0&609(24) &    2&817(10) &    0&495(23) &    0&81(8) & G\\
 & Core 48 &    0&79(7) & \multicolumn{2}{c}{--} &    0&402(18) &    3&684(11) &    0&477(22) &    6&0(6) & SA\\
 & Core 74 &    0&66(8) & \multicolumn{2}{c}{--} &    0&268(17) &    4&245(13) &    0&38(3) &    5&1(5) & G\\
 & Core 109 &    1&93(8) & \multicolumn{2}{c}{--} &    0&743(17) &    5&836(4) &    0&361(9) &    4&2(4) & G\\
\hline
C$^{34}$S~(3--2)
 & Core 14 &    0&12(3) & \multicolumn{2}{c}{--} &    0&034(4) &    3&488(17) &    0&27(4) &    1&70(17) & G\\
 & Core 20 &    0&15(5) & \multicolumn{2}{c}{--} &    0&064(11) &    3&59(3) &    0&40(8) &    0&25(3) & G\\
 & Core 109 &    0&18(6) & \multicolumn{2}{c}{--} &    0&064(8) &    5&82(3) &    0&340(00) &    0&189(19) & G\\
\hline
CS~(3--2)
 & Core 06 &    0&68(6) & \multicolumn{2}{c}{--} &    0&220(10) &    3&480(7) &    0&303(15) & \multicolumn{2}{c}{--} & G\\
 & Core 14 &    0&14(4) & \multicolumn{2}{c}{--} &    0&072(7) &    3&59(3) &    0&48(4) &   38&(4) & G\\
 & Core 20 &    0&67(6) & \multicolumn{2}{c}{--} &    0&179(8) &    3&502(8) &    0&2500(00) &    5&7(6) & G\\
 &  &    0&65(6) & \multicolumn{2}{c}{--} &    0&173(8) &    3&797(9) &    0&2500(00) & \multicolumn{2}{c}{--} & \\
 & Core 40 &    1&09(9) & \multicolumn{2}{c}{--} &    0&270(13) &    3&414(6) &    0&234(15) & \multicolumn{2}{c}{--} & G\\
 & Core 47 &    0&35(6) & \multicolumn{2}{c}{--} &    0&148(11) &    2&896(15) &    0&39(3) & \multicolumn{2}{c}{--} & G\\
 & Core 48 &    0&28(5) & \multicolumn{2}{c}{--} &    0&122(10) &    3&772(17) &    0&41(4) & \multicolumn{2}{c}{--} & G\\
 & Core 65 &    0&16(4) & \multicolumn{2}{c}{--} &    0&124(11) &    5&07(4) &    0&73(9) & \multicolumn{2}{c}{--} & G\\
 & Core 74 &    0&34(5) & \multicolumn{2}{c}{--} &    0&100(8) &    4&200(11) &    0&277(20) & \multicolumn{2}{c}{--} & G\\
 & Core 109 &    1&01(8) & \multicolumn{2}{c}{--} &    0&366(14) &    5&810(7) &    0&339(15) &    4&2(4) & G\\
\hline
\hline
\end{tabular}
\smallskip

Footnotes $^a$ to $^e$ as in Table~\ref{tab_param1}.\\

\label{tab_param2}
\end{table*}

\begin{table*}
\caption{
Line parameters$^a$ (Continuation).
}
\begin{tabular}{cc r@{.}l r@{.}l r@{.}l r@{.}l r@{.}l r@{.}l c }
\noalign{\smallskip}
\hline\hline\noalign{\smallskip}
\multicolumn{1}{c}{Molecular} &
& \multicolumn{2}{c}{$T_{\rm MB}$ $^b$} 
& \multicolumn{2}{c}{$A \times \tau$ $^c$} 
& \multicolumn{2}{c}{$\int T_{\rm MB}\mbox{d}\texttt{v}$ $^b$} 
& \multicolumn{2}{c}{$\texttt{v}_{\rm LSR}$} 
& \multicolumn{2}{c}{$\Delta\texttt{v}_{\rm LSR}$} 
\\
\multicolumn{1}{c}{transition}
& \multicolumn{1}{c}{Source} 
& \multicolumn{2}{c}{(K)} 
& \multicolumn{2}{c}{(K)} 
& \multicolumn{2}{c}{(K\, km\,s$^{-1}$)} 
& \multicolumn{2}{c}{(km\,s$^{-1}$)} 
& \multicolumn{2}{c}{(km\,s$^{-1}$)}
& \multicolumn{2}{c}{$\tau$ $^d$}
& \multicolumn{1}{c}{Profile$^e$} \\
\hline
C$^{18}$O~(1--0)
 & Core 06 &    2&61(6) & \multicolumn{2}{c}{--} &    1&137(13) &    3&5180(20) &    0&409(6) &    0&33(3) & G\\
 & Core 14 &    4&10(6) & \multicolumn{2}{c}{--} &    1&875(13) &    3&4890(10) &    0&430(4) &    0&84(8) & G\\
 & Core 20 &    2&97(6) & \multicolumn{2}{c}{--} &    0&798(10) &    3&6600(20) &    0&253(4) &    0&31(3) & G\\
 & Core 47 &    2&51(6) & \multicolumn{2}{c}{--} &    1&116(14) &    2&791(3) &    0&417(6) &    0&50(5) & G\\
 & Core 74 &    2&51(5) & \multicolumn{2}{c}{--} &    0&961(10) &    4&1920(20) &    0&360(5) &    0&59(6) & G\\
 & Core 109 &    2&42(5) & \multicolumn{2}{c}{--} &    0&991(11) &    5&7640(20) &    0&384(5) &    0&51(5) & G\\
\hline
$^{13}$CO~(1--0)
 & Core 06 &    7&78(6) & \multicolumn{2}{c}{--} &    4&98(11) &    3&536(5) &    0&601(6) &    1&83(18) & G\\
 &  &    1&96(6) & \multicolumn{2}{c}{--} &    1&44(11) &    4&133(24) &    0&69(3) & \multicolumn{2}{c}{--} & \\
 & Core 14 &    7&13(6) & \multicolumn{2}{c}{--} &    4&407(18) &    3&4280(10) &    0&581(4) &    4&7(5) & G\\
 &  &    4&13(6) & \multicolumn{2}{c}{--} &    2&639(4) &    3&8680(20) &    0&600(6) & \multicolumn{2}{c}{--} & \\
 & Core 20 &    9&12(6) & \multicolumn{2}{c}{--} &    5&903(16) &    3&7060(10) &    0&6080(20) &    1&71(17) & G\\
 & Core 47 &    5&99(5) & \multicolumn{2}{c}{--} &    3&647(12) &    2&7550(10) &    0&5720(20) &    2&8(3) & G\\
 &  &    4&27(5) & \multicolumn{2}{c}{--} &    3&912(14) &    3&2300(20) &    0&862(4) & \multicolumn{2}{c}{--} & \\
 & Core 74 &    5&41(5) & \multicolumn{2}{c}{--} &    3&598(21) &    4&2320(20) &    0&625(4) &    3&3(3) & G\\
 &  &    2&15(5) & \multicolumn{2}{c}{--} &    1&829(22) &    5&221(5) &    0&800(11) & \multicolumn{2}{c}{--} & \\
 & Core 109 &    5&68(6) & \multicolumn{2}{c}{--} &    3&577(18) &    5&7990(10) &    0&591(3) &    2&8(3) & G\\
 &  &    0&83(6) & \multicolumn{2}{c}{--} &    1&11(3) &    3&275(14) &    1&26(4) & \multicolumn{2}{c}{--} & \\
\hline
C$^{18}$O~(2--1)
 & Core 06 &    4&23(13) & \multicolumn{2}{c}{--} &    1&437(18) &    3&5240(20) &    0&319(5) &    1&11(11) & G\\
 & Core 14 &    3&52(23) & \multicolumn{2}{c}{--} &    1&76(4) &    3&522(5) &    0&469(13) &    0&94(9) & G\\
 & Core 20 &    3&26(25) & \multicolumn{2}{c}{--} &    1&07(4) &    3&712(5) &    0&308(12) &    0&52(5) & G\\
 & Core 40 &    3&6(3) & \multicolumn{2}{c}{--} &    1&33(4) &    3&323(5) &    0&349(12) & \multicolumn{2}{c}{--} & G\\
 & Core 47 &    2&61(22) & \multicolumn{2}{c}{--} &    1&22(4) &    2&779(6) &    0&440(15) &    0&52(5) & G\\
 & Core 48 &    4&05(15) & \multicolumn{2}{c}{--} &    1&287(19) &    3&6750(20) &    0&299(5) & \multicolumn{2}{c}{--} & G\\
 & Core 65 &    2&65(14) & \multicolumn{2}{c}{--} &    1&599(24) &    4&936(4) &    0&567(10) & \multicolumn{2}{c}{--} & G\\
 & Core 74 &    2&54(21) & \multicolumn{2}{c}{--} &    0&77(3) &    4&216(5) &    0&284(12) &    0&89(9) & G\\
 & Core 109 &    3&15(11) & \multicolumn{2}{c}{--} &    1&035(16) &    5&7820(20) &    0&309(5) &    0&86(9) & G\\
\hline
$^{13}$CO~(2--1)
 & Core 06 &    6&23(12) & \multicolumn{2}{c}{--} &    4&51(5) &    3&604(4) &    0&681(9) &    6&2(6) & G\\
 &  &    1&50(12) & \multicolumn{2}{c}{--} &    0&56(4) &    4&260(10) &    0&35(3) & \multicolumn{2}{c}{--} & \\
 & Core 14 &    5&66(12) & \multicolumn{2}{c}{--} &    2&88(5) &    3&378(4) &    0&477(4) &    5&2(5) & G\\
 &  &    5&05(12) & \multicolumn{2}{c}{--} &    2&88(7) &    3&819(5) &    0&536(11) & \multicolumn{2}{c}{--} & \\
 & Core 20 &    7&50(13) & \multicolumn{2}{c}{--} &    6&00(4) &    3&7100(20) &    0&751(5) &    2&9(3) & G\\
 & Core 47 &    6&01(12) & \multicolumn{2}{c}{--} &    5&05(8) &    3&005(6) &    0&790(9) &    2&9(3) & G\\
 &  &    3&08(12) & \multicolumn{2}{c}{--} &    0&95(7) &    2&631(5) &    0&291(12) & \multicolumn{2}{c}{--} & \\
 & Core 74 &    4&24(6) & \multicolumn{2}{c}{--} &    2&417(17) &    4&2550(20) &    0&536(4) &    4&9(5) & G\\
 &  &    1&79(6) & \multicolumn{2}{c}{--} &    1&168(19) &    5&259(5) &    0&614(12) & \multicolumn{2}{c}{--} & \\
 & Core 109 &    5&36(11) & \multicolumn{2}{c}{--} &    2&80(3) &    5&8310(20) &    0&491(5) &    4&8(5) & G\\
 &  &    0&61(11) & \multicolumn{2}{c}{--} &    0&70(4) &    3&38(3) &    1&09(8) & \multicolumn{2}{c}{--} & \\
\hline
\hline
\end{tabular}
\smallskip

Footnotes $^a$ to $^e$ as in Table~\ref{tab_param1}.\\

\label{tab_param3}
\end{table*}

\begin{table*}
\caption{
Molecular column densities of the chemical species observed toward the \pipe\ cores in cm$^{-2}$.
}
\begin{footnotesize}
\begin{tabular}{c cccccccc}
\noalign{\smallskip}
\hline\hline\noalign{\smallskip}
\multicolumn{1}{c}{Source}&
\multicolumn{1}{c}{C$_3$H$_2$~(2$_{1,2}$--1$_{1,0}$)$^a$}&
\multicolumn{1}{c}{C$_2$H~(1--0)}& 
\multicolumn{1}{c}{HCN~(1--0)}&
\multicolumn{1}{c}{N$_2$H$^+$~(1--0)}& 
\multicolumn{1}{c}{C$^{34}$S~(2--1)$^b$}&
\multicolumn{1}{c}{CH$_{3}$OH~(2$_{0,2}$--1$_{0,1}$)$^a$}& 
\multicolumn{1}{c}{CS~(2--1)$^b$}& 
\multicolumn{1}{c}{C$^{18}$O~(1--0)}\\
\noalign{\smallskip}
\hline\noalign{\smallskip}
Core 06 & $\phantom{< \ }6.40\times10^{11}$ & $\phantom{< \ }4.15\times10^{12}$ & $\phantom{< \ }2.22\times10^{11}$ & $\phantom{< \ }5.20\times10^{11}$ & $\phantom{< \ }2.95\times10^{11}$ & $\phantom{< \ }2.62\times10^{13}$ & $\phantom{< \ }1.23\times10^{13}$ & $\phantom{< \ }1.42\times10^{15}$\\
Core 14 & $\phantom{< \ }3.50\times10^{11}$ & $\phantom{< \ }3.05\times10^{12}$ & $<5.42\times10^{10}$ & $\phantom{< \ }9.70\times10^{10}$ & $\phantom{< \ }6.01\times10^{11}$ & $\phantom{< \ }1.64\times10^{13}$ & $\phantom{< \ }3.05\times10^{13}$ & $\phantom{< \ }2.78\times10^{15}$\\
Core 20 & $<1.78\times10^{11}$ & $\phantom{< \ }5.12\times10^{12}$ & $\phantom{< \ }3.08\times10^{11}$ & $<4.09\times10^{10}$ & $\phantom{< \ }3.05\times10^{11}$ & $\phantom{< \ }4.18\times10^{12}$ & $\phantom{< \ }1.18\times10^{13}$ & $\phantom{< \ }1.05\times10^{15}$\\
Core 40 & $\phantom{< \ }2.91\times10^{12}$ & $\phantom{< \ }1.22\times10^{13}$ & $\phantom{< \ }2.57\times10^{12}$ & $\phantom{< \ }4.89\times10^{11}$ & $\phantom{< \ }2.94\times10^{11}$ & $\phantom{< \ }1.28\times10^{13}$ & $\phantom{< \ }7.19\times10^{12}$& -- \\
Core 47 & $\phantom{< \ }3.52\times10^{11}$ & $\phantom{< \ }5.12\times10^{12}$& --  & $\phantom{< \ }1.30\times10^{11}$ & $\phantom{< \ }2.86\times10^{11}$ & $\phantom{< \ }4.54\times10^{12}$& --  & $\phantom{< \ }1.40\times10^{15}$\\
Core 48 & $<7.68\times10^{10}$ & $\phantom{< \ }3.85\times10^{12}$ & $\phantom{< \ }2.59\times10^{12}$ & $<3.79\times10^{10}$ & $\phantom{< \ }2.95\times10^{11}$ & $\phantom{< \ }2.23\times10^{12}$ & $\phantom{< \ }1.51\times10^{13}$& -- \\
Core 65 & $<7.25\times10^{10}$ & $<1.33\times10^{11}$& --  & $<3.68\times10^{10}$ & $<1.07\times10^{11}$ & $\phantom{< \ }3.64\times10^{12}$& -- & -- \\
Core 74 & $<1.37\times10^{11}$ & -- & $<4.76\times10^{10}$ & $<4.46\times10^{10}$ & $\phantom{< \ }2.20\times10^{11}$ & $\phantom{< \ }2.39\times10^{12}$ & $\phantom{< \ }9.88\times10^{12}$ & $\phantom{< \ }1.24\times10^{15}$\\
Core 109 & $\phantom{< \ }1.63\times10^{13}$ & $\phantom{< \ }1.05\times10^{13}$ & $\phantom{< \ }9.67\times10^{12}$ & $\phantom{< \ }6.79\times10^{11}$ & $\phantom{< \ }3.67\times10^{11}$ & $\phantom{< \ }1.36\times10^{13}$ & $\phantom{< \ }1.24\times10^{13}$ & $\phantom{< \ }1.24\times10^{15}$\\
\hline
\end{tabular}
\smallskip

$^a$ Transition with no opacity measurements available, thus optically thin
emission is assumed to obtain lower limits of the column densities.\\
$^b$ We assume optically thin emission for some cores with no data or no 
detection in \cs/\cts\ to obtain a lower limit of the column density.

\end{footnotesize}
\label{tab_mol_col_dens1}
\end{table*}

\begin{table*}
\caption{
Molecular column densities of the chemical species observed toward the \pipe\ 
cores in cm$^{-2}$ (continuation).
}
\begin{footnotesize}
\begin{tabular}{c cccccccc}
\noalign{\smallskip}
\hline\hline\noalign{\smallskip}
\multicolumn{1}{c}{Source}& 
\multicolumn{1}{c}{$^{13}$CO~(1--0)}&
\multicolumn{1}{c}{CN~(1--0)}& 
\multicolumn{1}{c}{C$^{34}$S~(3--2)$^a$}&
\multicolumn{1}{c}{CS~(3--2)$^a$}& 
\multicolumn{1}{c}{N$_2$D$^+$~(2--1)$^b$}&
\multicolumn{1}{c}{DCO$^+$~(3--2)$^b$}& 
\multicolumn{1}{c}{C$^{18}$O~(2--1)}& 
\multicolumn{1}{c}{$^{13}$CO~(2--1)}\\
\noalign{\smallskip}
\hline\noalign{\smallskip}
Core 06 & $\phantom{< \ }2.30\times10^{16}$ & $\phantom{< \ }1.20\times10^{12}$ & $<7.39\times10^{10}$ & $\phantom{< \ }4.49\times10^{11}$ & $<8.58\times10^{ 8}$ & $\phantom{< \ }7.39\times10^{11}$ & $\phantom{< \ }9.05\times10^{14}$ & $\phantom{< \ }2.12\times10^{16}$\\
Core 14 & $\phantom{< \ }4.14\times10^{16}$ & $\phantom{< \ }1.16\times10^{12}$ & $\phantom{< \ }1.12\times10^{12}$ & $\phantom{< \ }4.26\times10^{13}$ & $<7.91\times10^{ 9}$ & $<1.43\times10^{13}$ & $\phantom{< \ }1.04\times10^{15}$ & $\phantom{< \ }1.16\times10^{16}$\\
Core 20 & $\phantom{< \ }1.39\times10^{16}$ & $<1.58\times10^{11}$ & $\phantom{< \ }3.19\times10^{11}$ & $\phantom{< \ }8.71\times10^{12}$ & $<1.73\times10^{10}$ & $<3.54\times10^{13}$ & $\phantom{< \ }5.21\times10^{14}$ & $\phantom{< \ }6.96\times10^{15}$\\
Core 40& --  & $\phantom{< \ }4.69\times10^{12}$& --  & $\phantom{< \ }6.18\times10^{11}$ & $\phantom{< \ }3.63\times10^{ 9}$ & $<7.60\times10^{11}$ & $\phantom{< \ }9.33\times10^{14}$& -- \\
Core 47 & $\phantom{< \ }2.11\times10^{16}$ & $\phantom{< \ }2.49\times10^{12}$ & $<6.22\times10^{10}$ & $\phantom{< \ }3.02\times10^{11}$ & $<4.38\times10^{ 9}$& --  & $\phantom{< \ }5.96\times10^{14}$ & $\phantom{< \ }1.17\times10^{16}$\\
Core 48& --  & $<1.23\times10^{11}$ & $<9.17\times10^{10}$ & $\phantom{< \ }2.62\times10^{11}$ & $<3.02\times10^{ 9}$ & $<1.37\times10^{12}$ & $\phantom{< \ }9.77\times10^{14}$& -- \\
Core 65& --  & $<9.79\times10^{10}$ & $<6.79\times10^{10}$ & $\phantom{< \ }3.24\times10^{11}$ & $<3.09\times10^{ 9}$& --  & $\phantom{< \ }9.35\times10^{14}$& -- \\
Core 74 & $\phantom{< \ }2.36\times10^{16}$ & $<1.31\times10^{11}$ & $<1.18\times10^{11}$ & $\phantom{< \ }2.06\times10^{11}$ & $<3.76\times10^{ 9}$ & $<4.30\times10^{11}$ & $\phantom{< \ }4.66\times10^{14}$ & $\phantom{< \ }9.66\times10^{15}$\\
Core 109 & $\phantom{< \ }2.10\times10^{16}$ & $\phantom{< \ }2.82\times10^{12}$ & $\phantom{< \ }2.34\times10^{11}$ & $\phantom{< \ }5.09\times10^{12}$ & $\phantom{< \ }1.39\times10^{11}$ & $\phantom{< \ }1.13\times10^{12}$ & $\phantom{< \ }5.94\times10^{14}$ & $\phantom{< \ }1.04\times10^{16}$\\
\hline
\end{tabular}
\smallskip

$^a$ We assume optically thin emission for some cores with no data or no 
detection in \cts\ to obtain a lower limit of the column density.\\
$^b$ Transition with no opacity measurements available, thus optically thin emission
is assumed to obtain lower limits of the column densities.
\end{footnotesize}
\label{tab_mol_col_dens2}
\end{table*}

\begin{table*}[htbp] 

\caption{ H$_2$ column densities, $N_{\rm H_2}$, of the \pipe\ cores in
cm$^{-2}$~$^a$.}

\begin{center}
\begin{tabular}{c cccc c cc}
\noalign{\smallskip}
\hline\hline\noalign{\smallskip}
 & \multicolumn{4}{c}{Molecular survey$^b$} & & \multicolumn{2}{c}{CO survey$^c$}\\
\cline{2-5}
\cline{7-8}
\multicolumn{1}{c}{Source}  & 10.5$''$ & 15.0$''$ & 21.5$''$ & 27.0$''$ &  & 11.0$''$ & 22.5$''$\\
\noalign{\smallskip}
\hline\noalign{\smallskip}
Core 06 & $2.12\times10^{22}$ & $1.43\times10^{22}$ & $1.31\times10^{22}$ & $9.73\times10^{21}$ &  & $2.12\times10^{22}$ & $1.18\times10^{22}$\\
Core 14 & $1.59\times10^{22}$ & $1.26\times10^{22}$ & $1.10\times10^{22}$ & $1.01\times10^{22}$ &  & $7.48\times10^{21}$ & $7.79\times10^{21}$\\
Core 20 & $7.66\times10^{21}$ & $6.26\times10^{21}$ & $5.39\times10^{21}$ & $5.08\times10^{21}$ &  & $5.22\times10^{21}$ & $4.41\times10^{21}$\\
Core 40 & $1.32\times10^{22}$ & $1.28\times10^{22}$ & $1.12\times10^{22}$ & $1.07\times10^{22}$ &  & $1.36\times10^{22}$ & $1.06\times10^{22}$\\
Core 47 & $1.03\times10^{22}$ & $6.93\times10^{21}$ & $5.59\times10^{21}$ & $5.03\times10^{21}$ &  & $8.64\times10^{21}$ & $3.96\times10^{21}$\\
Core 48 & $1.09\times10^{22}$ & $8.88\times10^{21}$ & $7.38\times10^{21}$ & $6.99\times10^{21}$ &  & $1.34\times10^{22}$ & $6.71\times10^{21}$\\
Core 65 & $1.16\times10^{22}$ & $1.15\times10^{22}$ & $1.12\times10^{22}$ & $8.34\times10^{21}$ &  & $1.16\times10^{22}$ & $1.01\times10^{22}$\\
Core 74 & $6.98\times10^{21}$ & $6.75\times10^{21}$ & $5.96\times10^{21}$ & $5.65\times10^{21}$ &  & $1.16\times10^{22}$ & $5.74\times10^{21}$\\
Core 109 & $4.19\times10^{22}$ & $3.73\times10^{22}$ & $3.23\times10^{22}$ & $3.08\times10^{22}$ &  & $4.49\times10^{22}$ & $3.10\times10^{22}$\\
\hline
\end{tabular}
\end{center}

$^a$ Average column densities are calculated within one beam area. The values of
$\kappa_{\rm 250~GHz}$ and $T_{\rm dust}$ are the same as for
Table~\ref{tab_param_mambo}. These values are combined with the molecular column
densities to find the molecular abundances in the same beam area. \\
$^b$ Observations toward the dust continuum emission peak (Table~\ref{tab_param_mambo}). The correspondence is:
10.$\!''$5 with DCO$^+$, CN~(2--1), N$_2$D$^+$~(3--2) and H$^{13}$CO$^+$~(3--2);
15.$\!''$0 with C$^{34}$S~(3--2), CS~(3--2), and N$_2$D$^+$~(2--1);
21.$\!''$5 with CN~(1--0); and, finally, 
27.$\!''$0 with C$_3$H$_2$~(2--1), HCN~(1--0), N$_2$H$^+$~(1--0), C$^{34}$S~(2--1), CH$_3$OH~(2--1) and CS~(2--1).\\
$^c$ Observations toward the extinction peak (Table~\ref{tab_source}). The correspondence is:
11.$\!''$0 with C$^{18}$O~(2--1), and $^{13}$CO~(2--1);
22.$\!''$5 with C$^{18}$O~(1--0), and $^{13}$CO~(1--0).

\label{tab_dust_col_dens}
\end{table*}

\begin{table*}
\begin{footnotesize}
\caption{
Abundances$^a$ of the chemical species with respect to H$_2$ observed toward the
\pipe\ cores.
}
\begin{tabular}{c cccccccc}
\noalign{\smallskip}
\hline\hline\noalign{\smallskip}
\multicolumn{1}{c}{Source}&
\multicolumn{1}{c}{C$_3$H$_2$$^b$}&
\multicolumn{1}{c}{C$_2$H}& \multicolumn{1}{c}{HCN}&
\multicolumn{1}{c}{N$_2$H$^+$}& 
\multicolumn{1}{c}{C$^{34}$S$^c$}&
\multicolumn{1}{c}{CH$_{3}$OH$^b$}& 
\multicolumn{1}{c}{CS$^c$}& 
\multicolumn{1}{c}{C$^{18}$O}\\
\noalign{\smallskip}
\hline\noalign{\smallskip}
Core 06 & $\phantom{< \ }6.58\times10^{-11}$ & $\phantom{< \ }4.27\times10^{-10}$ & $\phantom{< \ }2.28\times10^{-11}$ & $\phantom{< \ }5.35\times10^{-11}$ 
	& $\phantom{< \ }3.03\times10^{-11}$ & $\phantom{< \ }2.69\times10^{-9}$ & $\phantom{< \ }1.27\times10^{-9}$ & $\phantom{< \ }1.21\times10^{-7}$\\
Core 14 & $\phantom{< \ }3.47\times10^{-11}$ & $\phantom{< \ }3.02\times10^{-10}$ & $<5.37\times10^{-12}$ & $\phantom{< \ }9.61\times10^{-12}$ & 
	$\phantom{< \ }5.95\times10^{-11}$ & $\phantom{< \ }1.63\times10^{-9}$ & $\phantom{< \ }3.02\times10^{-9}$ & $\phantom{< \ }3.57\times10^{-7}$\\
Core 20 & $<3.50\times10^{-11}$ & $\phantom{< \ }1.01\times10^{-9}$ & $\phantom{< \ }6.06\times10^{-11}$ & $<8.05\times10^{-12}$ 
	& $\phantom{< \ }6.01\times10^{-11}$ & $\phantom{< \ }8.23\times10^{-10}$ & $\phantom{< \ }2.32\times10^{-9}$ & $\phantom{< \ }2.38\times10^{-7}$\\
Core 40 & $\phantom{< \ }2.73\times10^{-10}$ & $\phantom{< \ }1.14\times10^{-9}$ & $\phantom{< \ }2.41\times10^{-10}$ & $\phantom{< \ }4.58\times10^{-11}$ 
	& $\phantom{< \ }2.76\times10^{-11}$ & $\phantom{< \ }1.20\times10^{-9}$ & $\phantom{< \ }6.74\times10^{-10}$& -- \\
Core 47 & $\phantom{< \ }6.99\times10^{-11}$ & $\phantom{< \ }1.02\times10^{-9}$& --  & $\phantom{< \ }2.58\times10^{-11}$ 
	& $\phantom{< \ }5.68\times10^{-11}$ & $\phantom{< \ }9.02\times10^{-10}$& --  & $\phantom{< \ }3.52\times10^{-7}$\\
Core 48 & $<1.10\times10^{-11}$ & $\phantom{< \ }5.51\times10^{-10}$ & $\phantom{< \ }3.71\times10^{-10}$ & $<5.43\times10^{-12}$ 
	& $\phantom{< \ }4.21\times10^{-11}$ & $\phantom{< \ }3.19\times10^{-10}$ & $\phantom{< \ }2.17\times10^{-9}$& -- \\
Core 65 & $<8.70\times10^{-12}$ & $<1.15\times10^{-11}$& --  & $<4.41\times10^{-12}$ 
	& $<9.29\times10^{-12}$ & $\phantom{< \ }4.36\times10^{-10}$& -- & -- \\
Core 74 & $<2.42\times10^{-11}$ & -- & $<8.43\times10^{-12}$ & $<7.90\times10^{-12}$ 
	& $\phantom{< \ }3.89\times10^{-11}$ & $\phantom{< \ }4.23\times10^{-10}$ & $\phantom{< \ }1.75\times10^{-9}$ & $\phantom{< \ }2.15\times10^{-7}$\\
Core 109 & $\phantom{< \ }5.28\times10^{-10}$ & $\phantom{< \ }3.41\times10^{-10}$ & $\phantom{< \ }3.14\times10^{-10}$ & $\phantom{< \ }2.20\times10^{-11}$ 
	& $\phantom{< \ }1.19\times10^{-11}$ & $\phantom{< \ }4.43\times10^{-10}$ & $\phantom{< \ }4.02\times10^{-10}$ & $\phantom{< \ }3.99\times10^{-8}$\\
\hline
\end{tabular}

$^a$ See Tables~\ref{tab_mol_col_dens1}, \ref{tab_mol_col_dens2}, and 
\ref{tab_dust_col_dens}
 for line and dust column densities.\\
$^b$ Transition with no opacity measurements available, thus optically thin emission is assumed to 
estimate a lower limit of the column densities and, consequently, of the abundances.\\
$^c$ We assume optically thin emission for some cores with no data or no 
detection in
\cs/\cts\ to obtain a lower limit 
of the column density and, as a result, also for the abundance.

\label{tab_abun1}
\end{footnotesize}

\end{table*}

\begin{table*}[h]
\begin{footnotesize}
\caption{
Abundances$^a$ of the chemical species with respect to H$_2$ observed toward the
\pipe\ cores.
}
\begin{tabular}{c cccccc}
\noalign{\smallskip}
\hline\hline\noalign{\smallskip}
\multicolumn{1}{c}{Source}& 
\multicolumn{1}{c}{$^{13}$CO}&
\multicolumn{1}{c}{CN}& 
\multicolumn{1}{c}{N$_2$D$^+$$^b$}&
\multicolumn{1}{c}{DCO$^+$$^b$}& 
\\
\noalign{\smallskip}
\hline\noalign{\smallskip}
Core 06 & $\phantom{< \ }1.95\times10^{-6}$ & $\phantom{< \ }9.16\times10^{-11}$ 
	& $<8.82\times10^{-14}$ & $\phantom{< \ }3.49\times10^{-11}$ \\ 
Core 14 & $\phantom{< \ }5.31\times10^{-6}$ & $\phantom{< \ }1.06\times10^{-10}$ 
	& $<7.84\times10^{-13}$ & $<1.83\times10^{-9}$  \\ 
Core 20 & $\phantom{< \ }3.15\times10^{-6}$ & $<3.10\times10^{-11}$ 
	& $<3.41\times10^{-12}$ & $<8.02\times10^{-9}$ \\ 
Core 40 & --  & $\phantom{< \ }4.19\times10^{-10}$
	& $\phantom{< \ }2.83\times10^{-13}$ & $<7.19\times10^{-11}$ \\ 
Core 47 & $\phantom{< \ }5.32\times10^{-6}$ & $\phantom{< \ }4.45\times10^{-10}$ 
	& $<8.70\times10^{-13}$& --  \\ 
Core 48 & --  & $<1.76\times10^{-11}$ 
	& $<4.33\times10^{-13}$ & $<2.04\times10^{-10}$ \\ 
Core 65 & --  & $<1.17\times10^{-11}$ 
	& $<3.71\times10^{-13}$& --  \\ 
Core 74 & $\phantom{< \ }4.11\times10^{-6}$ & $<2.33\times10^{-11}$ 
	& $<6.65\times10^{-13}$ & $<7.50\times10^{-11}$ \\ 
Core 109 & $\phantom{< \ }6.76\times10^{-7}$ & $\phantom{< \ }8.72\times10^{-11}$ 
	& $\phantom{< \ }3.73\times10^{-12}$ & $\phantom{< \ }2.69\times10^{-11}$ \\ 
\hline
\end{tabular}

$^a$ See Tables~\ref{tab_mol_col_dens1}, \ref{tab_mol_col_dens2}, and 
\ref{tab_dust_col_dens}
 for line and dust column densities.\\
$^b$ Transition with no opacity measurements available, thus optically thin emission is assumed to 
estimate a lower limit of the column densities and, consequently, of the abundances.\\
\label{tab_abun2}
\end{footnotesize}

\end{table*}

\twocolumngrid

   \begin{figure*}[t]
   \centering
   \includegraphics[height=\textwidth,angle=-90]{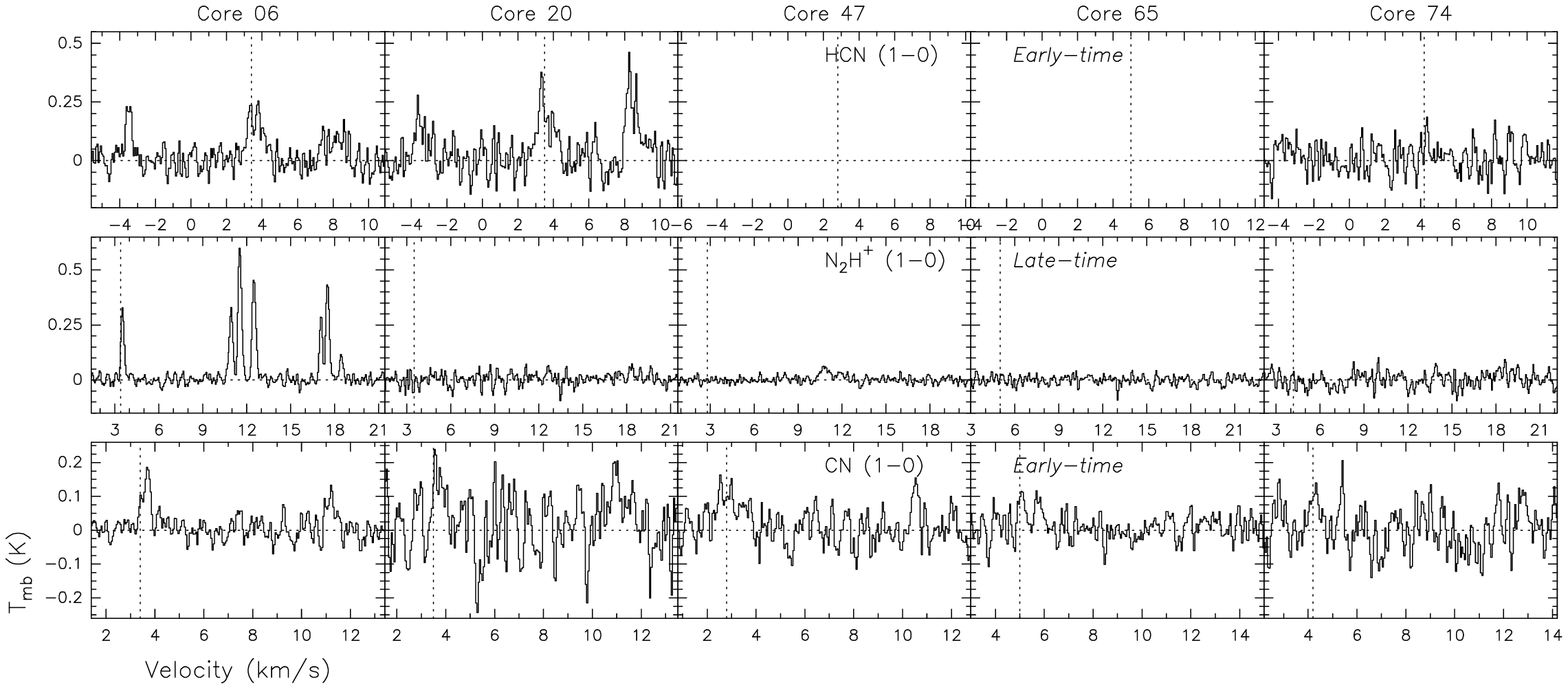}

      \caption{IRAM 30-m line spectra of the molecular transitions with
      hyperfine components presented in \paper\ toward the five new selected
      cores of the \pipe\ presented in this work (Table~\ref{tab_source}). 
      {\it Columns}: single cores named above the top panel of each column.
      {\it Rows}: single molecular transition specified on the third column.
      Empty panels represent non-observed molecular lines.
      The velocity range is 16.5, 20 and 12~km~s$^{-1}$ for \hcn~(1--0),
      \ndhp~(1--0), and \cn~(1--0), respectively. Horizontal axis shows the
      velocity, and the $\texttt{v}_{\rm LSR}$ of each core is marked with a
      vertical dotted line. Vertical axis shows the \Tmb\ of the emission, and
      the zero level is marked by a horizontal dotted line. 
      \label{fig:lineshfs} }

   \end{figure*}

   \begin{figure}[b]
   \includegraphics[width=\columnwidth,angle=0]{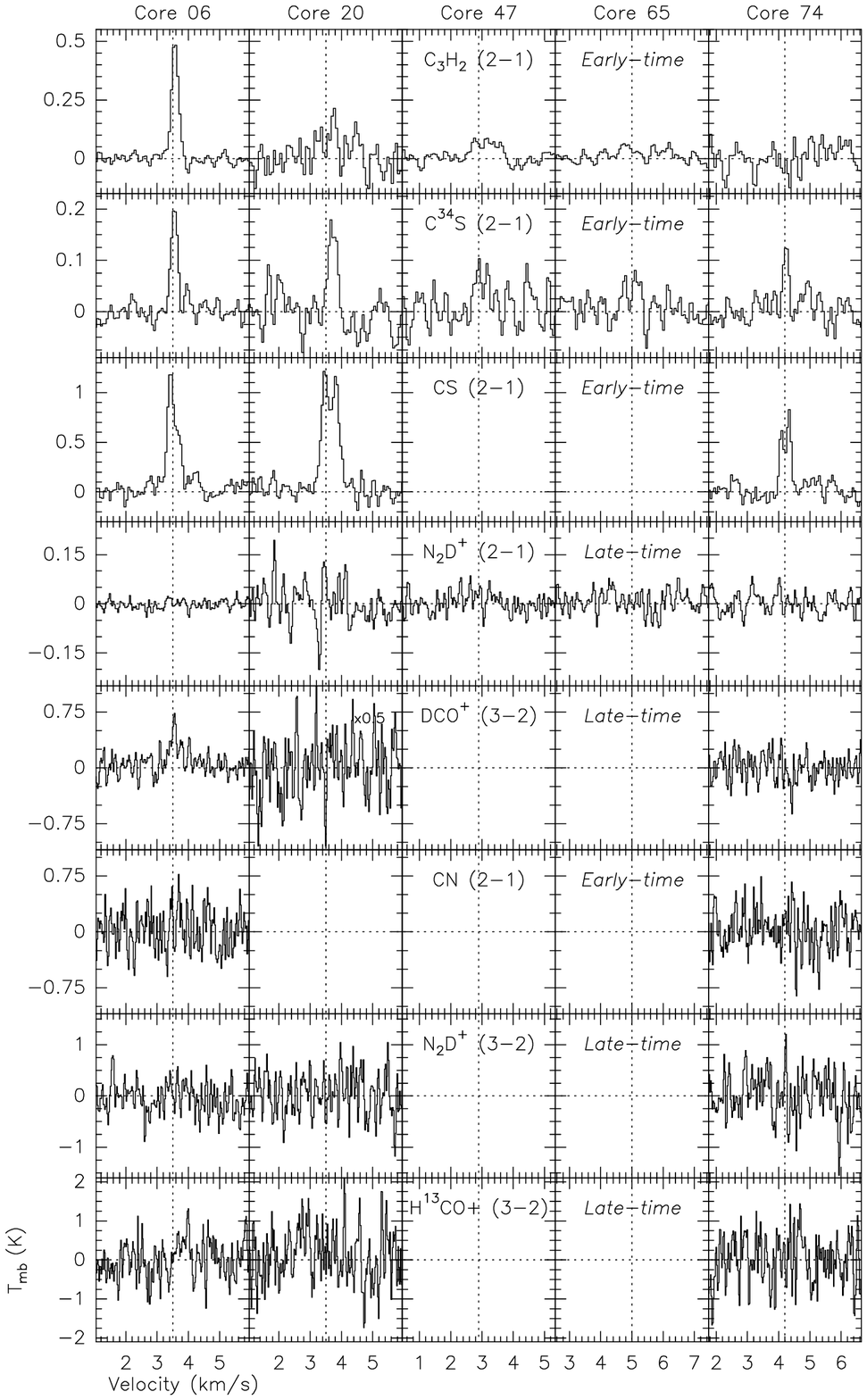}

      \caption{IRAM 30-m line spectra of the molecular transitions without
      hyperfine components presented in \paper\ toward the five new selected
      cores of the \pipe\ presented in this work (Table~\ref{tab_source}). {\it
      Columns}: single cores named above the top panel of each column. {\it
      Rows}: single molecular transition specified on the third column. 
      Empty panels represent non-observed molecular lines.
      Axes
      and dotted lines are as in Fig.~\ref{fig:lineshfs}. The velocity range is
      5~km~s$^{-1}$ centered on the $\texttt{v}_{\rm LSR}$ of each core. 
      \label{fig:lines1} }

   \end{figure}

   \begin{figure}
   \centering
   \includegraphics[width=\columnwidth,angle=0]{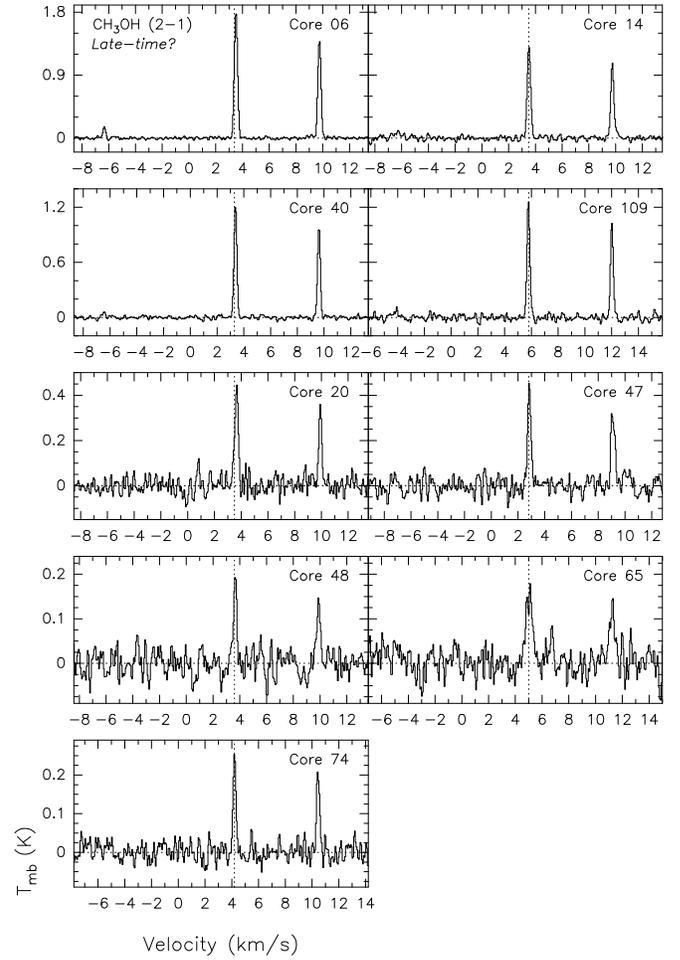}

      \caption{IRAM 30-m line spectra of \chtoh~(2--1) toward the nine selected
      cores of the \pipe\ (Table~\ref{tab_source}).  This molecular transition
      is not presented in \paper. The name of the core is indicated in the top
      right corner of each panel. Axes and dotted lines are as in
      Fig.~\ref{fig:lineshfs}. The velocity range is 22~km~s$^{-1}$.
      \label{fig:lineshfs_new} }

   \end{figure}

\clearpage
\onecolumngrid

   \begin{figure*}
   \centering
   \includegraphics[width=\textwidth,angle=0]{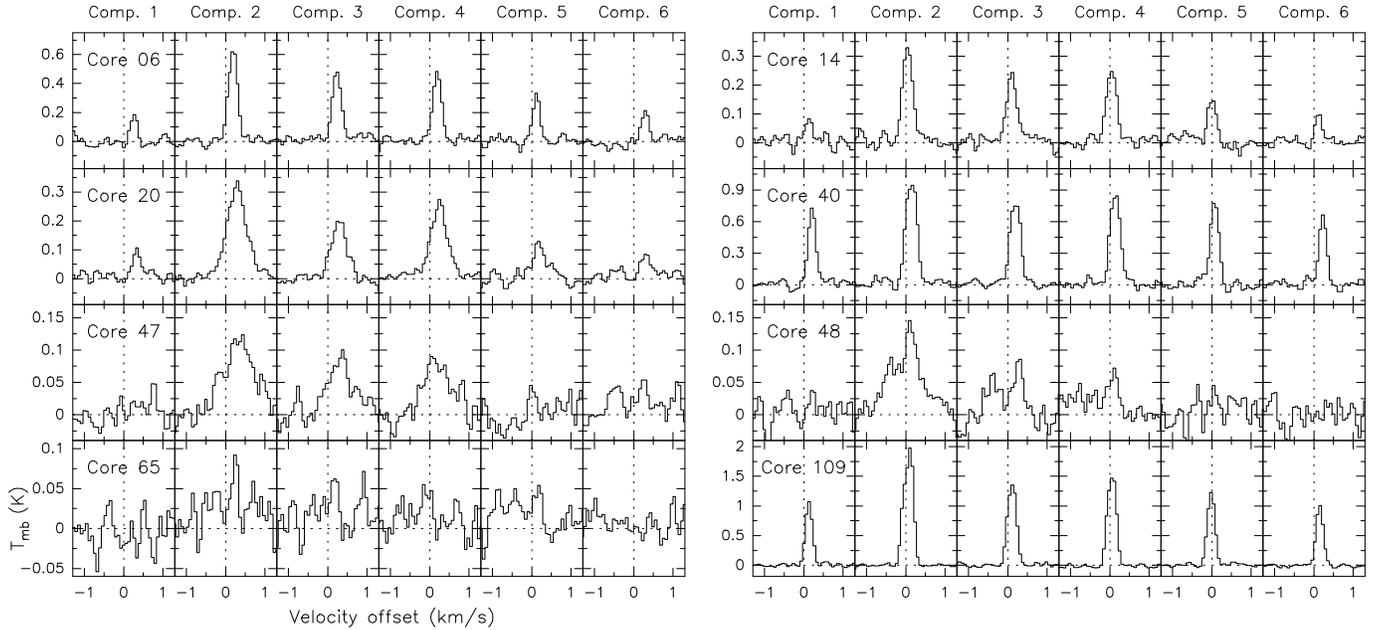}

      \caption{IRAM 30-m line spectra of \cdh~(1--0) (\textit{early-time}) toward
      eight cores of the \pipe\ (Table~\ref{tab_source}).  This
      molecular transition is not presented in \paper. The component number
      following \citet{padovani09} is indicated above each column. The name of the
      core is indicated in the left panel of each row. Axes and dotted lines are
      as in Fig.~\ref{fig:lineshfs}. The velocity range is 2.5~km~s$^{-1}$. 
      \label{fig:lineshfs_new2} }

   \end{figure*}

   \begin{figure*}
   \centering
   \includegraphics[width=.8\textwidth,angle=0]{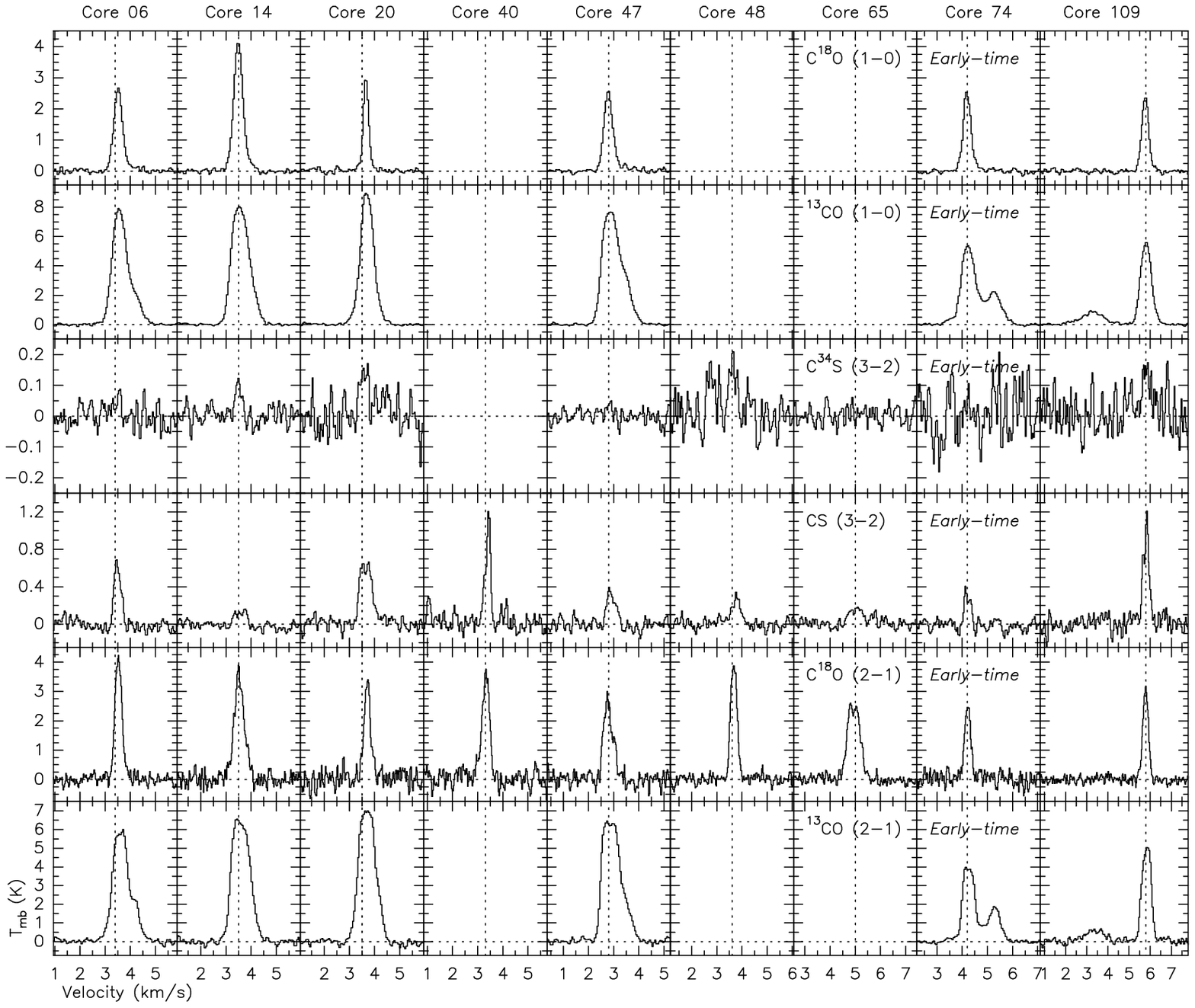}

      \caption{IRAM 30-m line spectra of the molecular transitions without
      hyperfine components toward the nine selected cores of the \pipe\
      (Table~\ref{tab_source}). These molecular transitions are not presented
      in \paper.  {\it Columns}: single cores named above the top panel of each
      column. {\it Rows}: single molecular transition specified on the seventh
      column. 
      Empty panels represent non-observed molecular lines.
      Axes and dotted lines are as in Fig.~\ref{fig:lineshfs}. The
      velocity range is 5~km~s$^{-1}$ except for Core~109 (6~km~s$^{-1}$).
        \label{fig:lines1_new} }

   \end{figure*}

\end{document}